\documentclass[journal]{IEEEtran}
\usepackage{cite}

\ifCLASSINFOpdf
\else
\fi
\usepackage[cmex10]{amsmath}
\usepackage{acronym} 
\usepackage{psfrag}   
\usepackage{bbm}
\usepackage{multirow,booktabs}
\usepackage{graphicx,subfig}
\usepackage{dsfont}

\begin{document}
%
\title{Robust Sequential Detection\\in Distributed Sensor Networks}
%
%
%

\author{Mark~R.~Leonard,~\IEEEmembership{Student Member,~IEEE,} 
        and~Abdelhak~M.~Zoubir,~\IEEEmembership{Fellow,~IEEE}
\thanks{M. R. Leonard and A. M. Zoubir are with the Signal Processing Group, Institute of Telecommunications, Technische Universit\"at Darmstadt, Darmstadt 64283, Germany (e-mail: leonard@spg.tu-darmstadt.de, zoubir@spg.tu-darmstadt.de)}
\thanks{Manuscript submitted }
}

%
%

\markboth{Submitted for publication}%
{}
%



\maketitle




%
\IEEEpeerreviewmaketitle

\begin{abstract}
We consider the problem of sequential binary hypothesis testing with a distributed sensor network in a non-Gaussian noise environment. To this end, we present a general formulation of the Consensus + Innovations Sequential Probability Ratio Test ($\mathcal{CI}$SPRT). Furthermore, we introduce two different concepts for robustifying the $\mathcal{CI}$SPRT and propose four different algorithms, namely, the Least-Favorable-Density-$\mathcal{CI}$SPRT, the Median-$\mathcal{CI}$SPRT, the M-$\mathcal{CI}$SPRT, and the Myriad-$\mathcal{CI}$SPRT. Subsequently, we analyze their suitability for different binary hypothesis tests before verifying and evaluating their performance in a shift-in-mean and a shift-in-variance scenario.
\end{abstract}
\section{Introduction}
\label{sec:intro}
The paradigm of sequential hypothesis testing is to make a reliable decision for one out of two or more hypotheses based on as few measurements as possible. This work is concerned with robust sequential binary hypothesis tests of the shift-in-mean and shift-in-variance type in distributed sensor networks. These kinds of tests play an important role in modern real-time applications such as intelligent traffic control, smart homes, or video surveillance \cite{Tartakovsky2014}. We consider sequential tests to make a decision as soon as enough data has been collected to guarantee a certain level of confidence \cite{wald1947sequential}. In addition, the tests are performed in a distributed fashion to avoid having a single point of failure and exploit the inherent scalability and fault-tolerance a sensor network provides \cite{akyildiz2002a-survey}. Furthermore, we are interested in tests that are robust against distributional uncertainties such as outliers encountered in real-world scenarios where the assumption of Gaussianity is often violated \cite{zoubir2012robust}.

The concepts of sequential detection \cite{wald1947sequential,novikov2008optimal,novikov2009optimal,Tartakovsky2014,fauss2015a-linear}, distributed signal processing in wireless sensor networks \cite{cattivelli2011distributed,tu2011mobile,sayed2012diffusion,balthasar2014,leonard2015distributed,matta2016distributed}, and robustness \cite{huber1964robustEst,huber1965robust,huber1973minimax, huber1981robust,levy2008principles,zoubir2012robust,gul2017minimax} are well-established fields of research. Also the combinations of either two of them, i.e., distributed sequential detection \cite{teneketzis1987decentralized,blum1997distributed,sahu2014distributed,sahu2016distributed}, robust sequential detection \cite{degroot1960minimax,schmitz1987minimax,fauss2015old-bands,gul2017minimax}, and robust hypothesis testing in distributed sensor networks \cite{veeravalli1994minimax,blum1997distributed,gul2017robust,gul2017theoretical,alsayed2017robust}, have received considerable attention in recent years. To the best of our knowledge, the union of sequential hypothesis testing, a distributed network architecture, and robustness is an important niche area that has not been treated in the literature, yet. 

This work contains the following contributions: We develop a general formulation for the Consensus + Innovations Sequential Probability Ratio Test ($\mathcal{CI}$SPRT) from \cite{sahu2016distributed} that (a) is applicable to arbitrary binary hypothesis tests, and (b) works with right-stochastic weighting matrices, which makes it suitable for common distributed detection problems. Furthermore, we propose a robust version of the $\mathcal{CI}$SPRT, dubbed LFD-$\mathcal{CI}$SPRT, based on the concept of least-favorable densities (LFDs). To this end, we derive the probability density function of the robust log-likelihood ratio of the LFDs and show how to calculate its mean and variance. Subsequently, we present an alternative method to robustify the $\mathcal{CI}$SPRT, which introduces robust estimators into the test statistic update. With the Median-$\mathcal{CI}$SPRT, the M-$\mathcal{CI}$SPRT, and the Myriad-$\mathcal{CI}$SPRT we propose three robust distributed sequential detectors based on this concept and analyze their suitability for Gaussian shift-in-mean and shift-in-variance tests. These results generalize and extend our first approaches from \cite{leonard2017robust} and \cite{hou2017robust} and provide a unified framework for robust sequential detection in distributed sensor networks. An extension of some of the presented concepts to multiple hypothesis tests can be found in \cite{leonard2018robust}.

The paper is structured as follows: In Section~\ref{sec:probform} we formulate the problem of distributed sequential shift-in-mean and shift-in-variance tests in the face of distributional uncentainties. Section~\ref{sec:cisprt} is dedicated to the reformulation of the $\mathcal{CI}$SPRT with the decision thresholds being derived in Section~\ref{sec:dec_th}. The robustification of the $\mathcal{CI}$SPRT based on LFDs is presented in Section~\ref{sec:lfd-cisprt}, the concept based on robust estimators in Section~\ref{sec:rob_est}. In Section~\ref{sec:simulations} we present simulation results, which verify and evaluate the performance of our proposed algorithms in a shift-in-mean and a shift-in-variance test. Conclusions are drawn in Section~\ref{sec:conclusion}.

\section{Problem Formulation}
\label{sec:probform}

Let $\boldsymbol{Y}(t) \in \mathds{R}^N, t=1,2,\ldots$ be a sequence of  random vectors with entries $Y_k(t), k=1,\ldots,N$. For all $k$ and $t$ the random variables $Y_k(t)$ are assumed to be independent and identically distributed according to distribution $P$, which admits a density $p$. Furthermore, consider a network of $N$ agents, which can be modeled as a simple, connected, and undirected graph $\mathcal{G} = (\mathcal{V},\mathcal{E})$ with $\mathcal{V}$ denoting the set of agents and $\mathcal{E}$ being the set of edges between these agents. The open neighborhood of agent $k$ is given by $\mathcal{N}_k = \{l \in \mathcal{V}\; |\; (k,l) \in \mathcal{E}\}$, i.e., the set of all agents to which $k$ is connected by an edge.

In distributed sequential detection, each agent $k$, $ k=1,\ldots,N$, sequentially performs a binary statistical hypothesis test to decide between the null hypothesis $\mathcal{H}_0$ and the alternative $\mathcal{H}_1$ given by
\begin{align*}
	\mathcal{H}_0 &: P = P_0\\
	\mathcal{H}_1 &: P = P_1.
\end{align*}
To this end, each node $k$ takes a measurement $y_k(t)$ at time instant $t$ from which a test statistic is computed. Considering a Gaussian environment, the hypotheses can be rewritten in terms of the corresponding random variable $Y_k(t)$ as
\begin{align}
	\begin{aligned}
	\mathcal{H}_0 &: Y_k(t) \sim \mathcal{N}(\mu_0,\sigma_0^2)\\	
	\mathcal{H}_1 &: Y_k(t) \sim\mathcal{N}(\mu_1,\sigma_1^2),
	\end{aligned}
	\label{eq:hyp_1}
\end{align}
where $\mu_i, i\in \{0,1\}$ is the known mean and $\sigma_i^2$ the variance of a zero-mean Gaussian noise process. While all of our results can be applied to any binary hypothesis test of this type, we will focus on the following two test scenarios:
\begin{enumerate}
	\item \emph{Scenario 1:} Shift-in-Mean Test \\
	We test for the mean of the distribution under the true hypothesis assuming equal variance $\sigma^2$ under both $\mathcal{H}_0$ and $\mathcal{H}_1$. The hypotheses become
	\begin{align}
	\begin{aligned}
	\mathcal{H}_0 &: Y_k(t)\sim\mathcal{N}(\mu_0,\sigma^2),\\
	\mathcal{H}_1 &: Y_k(t)\sim\mathcal{N}(\mu_1,\sigma^2).
	\end{aligned}
\end{align}
\item \emph{Scenario 2:} Shift-in-Variance Test \\
	We test for the variance of the distribution under the true hypothesis assuming two zero-mean Gaussian distributions. An example for this is a test for the presence or absence of a signal with known variance $\sigma_x^2$ in noise with power $\sigma_n^2$, i.e.,
	\begin{align}
	\begin{aligned}
	\mathcal{H}_0 &: Y_k(t)\sim\mathcal{N}(0,\sigma_n^2),\\
	\mathcal{H}_1 &: Y_k(t)\sim\mathcal{N}(0,\sigma_x^2+\sigma_n^2).
	\end{aligned}
\end{align}

\end{enumerate}

In many practical applications there is an uncertainty about the distribution of the data such that the assumption of Gaussian measurement noise might be violated. Taking these uncertainties into account transforms the test into one between two disjoint probability sets $\mathcal{P}_0$ and $\mathcal{P}_1$ with
\begin{align*}
	\mathcal{H}_0 &: P \in \mathcal{P}_0,\\
	\mathcal{H}_1 &: P \in \mathcal{P}_1.
\end{align*}
In Sections~\ref{sec:lfd-cisprt} and \ref{sec:rob_est} we present two different methods to robustify distributed sequential detectors based on the $\mathcal{CI}$SPRT against distributional uncertainties. We do this by limiting the influence of outliers on the test such that the required probabilities of false alarm and misdetection are still fulfilled. As the simulation results in Section~\ref{sec:simulations} show, this comes at no or minimal cost in terms of the average run length of the test.

\section{A General Formulation of the $\boldsymbol{\mathcal{CI}}$SPRT}
\label{sec:cisprt}
In \cite{sahu2014distributed,sahu2016distributed}, the authors propose the $\mathcal{CI}$SPRT as a distributed sequential detector based on the \emph{consensus+innovations} approach \cite{kar2013consensus}. In analogy to the centralized SPRT introduced by Wald \cite{wald1947sequential}, each agent $k$ in the $\mathcal{CI}$SPRT compares its test statistic $S_k(t)$ at time instant $t$ with an upper and a lower threshold to either decide for one of the two hypotheses if the respective threshold is crossed, or continue the test. $S_k(t)$ is recursively calculated as \cite{sahu2014distributed,sahu2016distributed}

\begin{align}
	S_k(t) &= \sum_{l\in \mathcal{N}_k\cup\{k\}}w_{kl}\left(S_l(t-1)+ \eta_l(t)\right),
	\label{eq:sk}
\end{align}
with $w_{kl}$ denoting appropriate combination weights that sum to one. Furthermore, $\eta_l(t)$ is the log-likelihood ratio of node $l$ at time instant $t$, which is calculated as

\begin{align}
	\begin{aligned}
		\eta_l(t) &= \log\!\left(\frac{p_1(y_l(t))}{p_0(y_l(t))}\right)\\
		&= \log\!\left(\frac{\sigma_0}{\sigma_1} e^{-\frac{(y_l(t)-\mu_1)^2}{2\sigma_1^2}} e^{\frac{(y_l(t)-\mu_0)^2}{2\sigma_0^2}}\right)\\
		&= \frac{\sigma_1^2\left(y_l(t)-\mu_0\right)^2-\sigma_0^2\left(y_l(t)-\mu_1\right)^2}{2\sigma_0^2\sigma_1^2} + \log\!\left(\frac{\sigma_0}{\sigma_1}\right)
	\end{aligned}
	\label{eq:eta_k}
\end{align}
assuming the general formulation from Eq.~\eqref{eq:hyp_1}. By collecting the combination weights into an $N \times N$ combination matrix, Eq.~\eqref{eq:sk} can be rewritten as
\begin{align}
	S_k(t) = \sum_{j=1}^t \boldsymbol{e}_k^\top\boldsymbol{W}^{t+1-l}\boldsymbol{\eta}(j),
	\label{eq:sk_vector}
\end{align}
with $\boldsymbol{e}_k$ denoting the $k$th column of identity matrix $\boldsymbol{I}$ of size $N$. Furthermore, vector $\boldsymbol{\eta}(j) = \left[\eta_1(j),\ldots,\eta_N(j)\right]^\top$ collects the log-likelihood ratios of all agents at time instant $j$.
In the sequel, we discuss how to choose the combination matrix $\boldsymbol{W}$.

\subsection{The Choice of Weighting Matrix $\boldsymbol{W}$}
\label{sec:w}
In \cite{sahu2016distributed}, the authors assume a weighting matrix that is non-negative, symmetric, irreducible, and stochastic by design. However, the design process relies on a method originally introduced in \cite{xiao2004fast}, which can and most of the time does produce a matrix with negative weights as explicitly stated by the authors. In the context of distributed detection, such a matrix is not practical since it will cause the information of some of the collaborating nodes to be given a negative weight. This operation has no meaning in distributed sensor networks.

Instead of requiring the weighting matrix to be non-negative, symmetric, irreducible, and stochastic, we consider a right-stochastic matrix, the rows of which sum up to one. Matrices of this kind are common, e.g., in the context of diffusion adaptation \cite{sayed2012diffusion}. An example for a right-stochastic matrix is one that puts equal weight on the information of the closed neighborhood of a node, i.e., the entries of $\boldsymbol{W}$ are given by
\begin{align*}
	w_{k,l} = \begin{cases}
 	\frac{1}{|\mathcal{N}_k\cup\{k\}|}&, l \in \mathcal{N}_k\cup\{k\}\\
 	0&, \text{otherwise}
 \end{cases}.
\end{align*}

\section{Decision Thresholds for the $\boldsymbol{\mathcal{CI}}SPRT$}
\label{sec:dec_th}
The decision thresholds derived in \cite{sahu2016distributed} suffer from two disadvantages. First, they only hold for the specific case of symmetric Gaussian shift-in-mean hypothesis tests. In \cite{leonard2017robust} and \cite{hou2017robust}, we generalized these thresholds for use in arbitrary binary hypothesis tests. Second, the derivation of the thresholds relies on the symmetry of $\boldsymbol{W}$, an assumption that is usually not valid in distributed sensor networks. In the sequel, we improve the generalized thresholds from \cite{leonard2017robust} and \cite{hou2017robust} by requiring only the right-stochasticity of $\boldsymbol{W}$ in the derivation. First, however, expressions for the mean and the variance of the test statistic under $\mathcal{H}_0$ and $\mathcal{H}_1$ are derived, which will be needed in the subsequent steps. 

\subsection{Mean and Variance of the Test Statistic}
\label{sec:mu_var_sk}

The expected value of the test statistic in Eq.~\eqref{eq:sk} under hypothesis $\mathcal{H}_i, i\in\{0,1\}$ is given by
\begin{align}
\begin{aligned}
	E_i\{S_k(t)\} &= \sum_{j=1}^t \boldsymbol{e}_k^\top\boldsymbol{W}^{t+1-j}E_i\{\boldsymbol{\eta}(j)\}\\	
	&= \mu_{\eta,i} \sum_{j=1}^t\boldsymbol{e}_k^\top\boldsymbol{W}^{t+1-j}\boldsymbol{1}\\
	&= \mu_{\eta,i} t,
	\end{aligned}
	\label{eq:mu_sk}
\end{align}
where $E_i\{\cdot\}$ denotes taking the expectation under hypothesis $\mathcal{H}_i$, $\mu_{\eta,i} = E_i\{\boldsymbol{\eta}(j)\}$ is the expected value of the log-likelihood ratio under $\mathcal{H}_i$, and $\boldsymbol{1}$ is the one-vector of length $N$. 

The variance of the test statistic in Eq.~\eqref{eq:sk} under hypothesis $\mathcal{H}_i$ can be calculated as
\begin{align*}
\begin{aligned}
	&\text{Var}_i\{S_k(t)\} = E_i\{(S_k(t)-\mu_{\eta,i} t)^2\}\\
	&= E_i\{S_k(t)^2\} - 2\mu_{\eta,i} t E_i\{S_k(t)^2\} + \mu_{\eta,i}^2t^2\\
	&= E_i\{S_k(t)^2\} - \mu_{\eta,i}^2t^2\\
	&= E_i\left\{\left(\sum_{j=1}^t \boldsymbol{e}_k^\top\boldsymbol{W}^{t+1-j}\boldsymbol{\eta}(j)\right)^2\right\} - \mu_{\eta,i}^2t^2\\
	&= \sum_{j=1}^t \sum_{l =1}^t \boldsymbol{e}_k^\top\boldsymbol{W}^{t+1-j} E_i\left\{\boldsymbol{\eta}(j)\boldsymbol{\eta}(l)\right\}\left(\boldsymbol{W}^{t+1-l}\right)^\top\boldsymbol{e}_k - \mu_{\eta,i}^2t^2.
	\end{aligned}
	\end{align*}
Since 
	\begin{align*}
	E_i\left\{\boldsymbol{\eta}(j)\boldsymbol{\eta}(l)\right\} = \begin{cases}
		\mu_{\eta,i}^2\boldsymbol{1}\boldsymbol{1}^\top &, j\neq l\\
		\sigma^2_{\eta,i}\boldsymbol{I} + \mu_{\eta,i}^2\boldsymbol{1}\boldsymbol{1}^\top &, j=l
	\end{cases},
	\end{align*}
	where $\sigma^2_{\eta,i}$ denotes the variance of the log-likelihood ratio under $\mathcal{H}_i$. By rearranging the two sums, we obtain

\begin{align*}
\begin{aligned}	
	\text{Var}_i\{S_k(t)\} &= \sigma^2_{\eta,i} \sum_{j=1}^t \boldsymbol{e}_k^\top\boldsymbol{W}^j \left(\boldsymbol{W}^j\right)^\top\boldsymbol{e}_k\\& \quad+ \mu_{\eta,i}^2 \sum_{j=1}^t \sum_{l=1}^t \boldsymbol{W}^j \boldsymbol{1}\boldsymbol{1}^\top\left(\boldsymbol{W}^l\right)^\top\boldsymbol{e}_k - \mu_{\eta,i}^2t^2\\
	= \sigma^2_{\eta,i} \sum_{j=1}^t& \boldsymbol{e}_k^\top\boldsymbol{W}^j \left(\boldsymbol{W}^j\right)^\top\boldsymbol{e}_k + \mu_{\eta,i}^2t^2 - \mu_{\eta,i}^2t^2\\
	= \sigma^2_{\eta,i} \sum_{j=1}^t& \boldsymbol{e}_k^\top\boldsymbol{W}^j \left(\boldsymbol{W}^j\right)^\top\boldsymbol{e}_k\\
	= \sigma^2_{\eta,i} \sum_{j=1}^t& \boldsymbol{e}_k^\top\boldsymbol{W}^{j-m}\boldsymbol{W}^m \left(\boldsymbol{W}^\top\right)^m\left(\boldsymbol{W}^\top\right)^{j-m}\boldsymbol{e}_k,
	\end{aligned}
\end{align*}
with $1\leq m\leq j$. By upper-bounding the $(k,k)$th entry of $\boldsymbol{W}^m \left(\boldsymbol{W}^\top\right)^m$ with a scalar $\xi$ according to
\begin{align*}
		\boldsymbol{e}_k\boldsymbol{W}^m \left(\boldsymbol{W}^\top\right)^m \boldsymbol{e}_k^\top \leq \xi \boldsymbol{1}\boldsymbol{1}^\top
\end{align*}
and using the properties 
\begin{align*}
	\begin{aligned}
	\boldsymbol{W}\boldsymbol{1}\boldsymbol{1}^\top = \boldsymbol{1}\boldsymbol{1}^\top\\
	\boldsymbol{1}\boldsymbol{1}^\top\boldsymbol{W}^\top = \boldsymbol{1}\boldsymbol{1}^\top\\
	\boldsymbol{W}\boldsymbol{1}\boldsymbol{1}^\top\boldsymbol{W}^\top = \boldsymbol{1}\boldsymbol{1}^\top
	\end{aligned}
\end{align*}
an upper bound on the variance of the test statistic can be found as
\begin{align}
	\text{Var}_i\{S_k(t)\} &\leq \sigma^2_{\eta,i} \xi t.
	\label{eq:var_sk}
\end{align}
A suitable choice for $\xi$ is the maximum value of the matrix $\boldsymbol{W}^m \left(\boldsymbol{W}^\top\right)^m$, i.e.,
\begin{align}
		\xi = \left\|\boldsymbol{W}^m \left(\boldsymbol{W}^\top\right)^m\right\|_\text{max},
\end{align}
where $\|\cdot\|_\text{max}$ is the max norm of a matrix. Another choice for $\xi$ is the largest eigenvalue of $\boldsymbol{W}^m \left(\boldsymbol{W}^\top\right)^m$ divided by the number of nodes $N$, i.e.,
\begin{align}
		\xi = \frac{1}{N}\lambda_\text{max}\!\left(\boldsymbol{W}^m \left(\boldsymbol{W}^\top\right)^m\right).
\end{align}
Note that the accuracy of this approximation can be tuned by the choice of $m$ and thus traded off against computational load. In most distributed sensor networks, computational power at the individual nodes is a scarce resource, which is why it makes sense to choose $m=1$. However, if more computational power is available, a higher accuracy can be achieved by choosing a larger value for $m$. 

The resulting expressions for the mean and the variance of the test statistic depend on the mean and the variance of the log-likelihood ratio $\eta_k(t)$ of node $k$ at time instant $t$. For a general binary hypothesis test as defined in Eq.~\eqref{eq:hyp_1} it can be shown that these quantities are given by 
\begin{align}
&\begin{aligned}
	\mu_{\eta,0} &= -\frac{\mu_0^2+\mu_1^2-2\mu_0\mu_1+\sigma_0^2-\sigma_1^2}{2\sigma_1^2} + \log\!\left(\frac{\sigma_0}{\sigma_1}\right)\\
	\mu_{\eta,1} &= \frac{\mu_0^2+\mu_1^2-2\mu_0\mu_1+\sigma_1^2-\sigma_0^2}{2\sigma_0^2} + \log\!\left(\frac{\sigma_0}{\sigma_1}\right)
	\label{eq:mu_eta}
	\end{aligned}\\
	&\begin{aligned}
	\sigma_{\eta,0}^2 &= \frac{1}{2}\left(1+\frac{\sigma_0^4}{\sigma_1^4}\right) + \left(\mu_0-\mu_1\right)^2 \frac{\sigma_0^2}{\sigma_1^4} - \frac{\sigma_0^2}{\sigma_1^2}\\
	\sigma_{\eta,1}^2 &= \frac{1}{2}\left(1+\frac{\sigma_1^4}{\sigma_0^4}\right) + \left(\mu_0-\mu_1\right)^2 \frac{\sigma_1^2}{\sigma_0^4} - \frac{\sigma_1^2}{\sigma_0^2}.	
	\label{eq:var_eta}
	\end{aligned}
\end{align}
The derivation of Eqs.~\eqref{eq:mu_eta} and \eqref{eq:var_eta} is detailed in Appendix \ref{app:mu_var_eta}.


\begin{figure*}[!t]
\psfrag{x1}[tc][tc][0.9]{$t=1$}
\psfrag{x2}[tc][tc][0.9]{$t=2$}
\psfrag{x3}[tc][tc][0.9]{$t=3$}
\psfrag{y1}[bc][bc][0.9]{probability density}
\psfrag{y}[bc][bc][0.9]{probability density}
\psfrag{x}[bc][bc][0.9]{}
\psfrag{y2}[bc][bc][0.9]{}
\psfrag{y3}[bc][bc][0.9]{}
\psfrag{-2}[tc][tc][0.9]{$-2$}
\psfrag{-1}[tc][tc][0.9]{$-1$}
\psfrag{0}[tc][tc][0.9]{$0$}
\psfrag{1}[tc][tc][0.9]{$1$}
\psfrag{2}[tc][tc][0.9]{$2$}
\psfrag{3}[tc][tc][0.9]{$3$}
\psfrag{4}[tc][tc][0.9]{$4$}
\psfrag{5}[tc][tc][0.9]{$5$}
\psfrag{-5}[tc][tc][0.9]{-$5$}
\psfrag{-3}[tc][tc][0.9]{-$3$}
\psfrag{6}[tc][tc][0.9]{$6$}
\psfrag{8}[tc][tc][0.9]{$8$}
\psfrag{10}[tc][tc][0.9]{$10$}
\psfrag{20}[tc][tc][0.9]{$20$}
\psfrag{-20}[tc][tc][0.9]{$-20$}
\psfrag{-40}[tc][tc][0.9]{$-40$}
\psfrag{-60}[tc][tc][0.9]{$-60$}
\psfrag{40}[tc][tc][0.9]{$40$}
\psfrag{0.5}[tc][tc][0.9]{$0.5$}
\psfrag{0.1}[tc][tc][0.9]{$0.1$}
\psfrag{-0.2}[tc][tc][0.9]{$-0.2$}
\psfrag{-0.4}[tc][tc][0.9]{$-0.4$}
\psfrag{-0.6}[tc][tc][0.9]{$-0.6$}
\psfrag{-0.8}[tc][tc][0.9]{$-0.8$}
\psfrag{1.5}[tc][tc][0.9]{$1.5$}
\psfrag{2.5}[tc][tc][0.9]{$2.5$}
\subfloat[]{
\includegraphics[height=4.4cm]{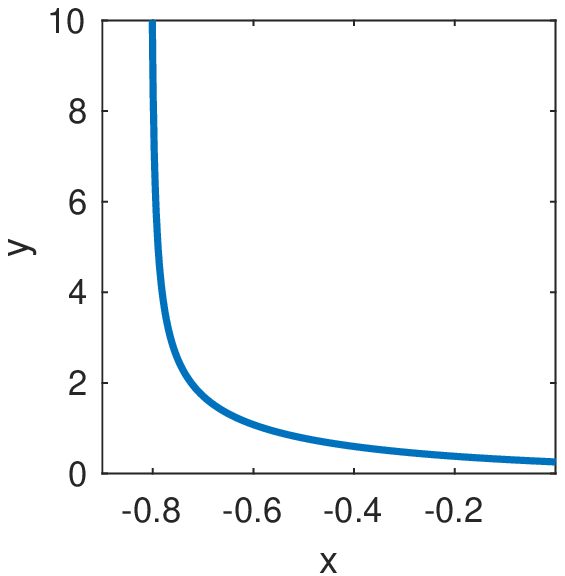}
\label{fig:chi2pdf}
}\hspace{-2.5em}
\subfloat[]{
\includegraphics[height=4.8cm]{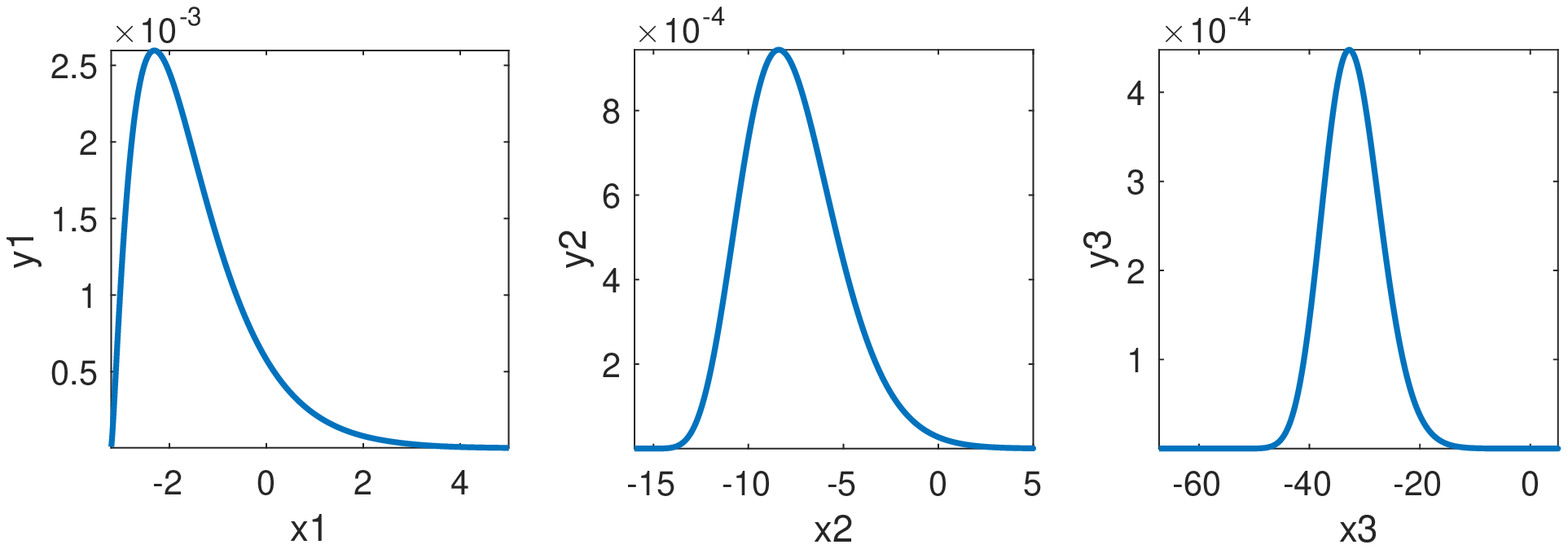}
\label{fig:vartest_pdf}
}
\caption{Shift-in-variance test: (a) probability density function of the log-likelihood ratio and (b) evolution over time of the probability density function of the test statistic $S_k(t)$ of an agent with 3 neighbors}	
\label{fig:vartest}
\end{figure*}

\subsection{Derivation of the Decision Thresholds}
\label{sec:deriv_th}
The test can easily be shown to terminate almost surely at a finite stopping time $T_k$ with
\begin{align*}
	T_k &= \inf\left\{ t | S_k(t) \notin \left[\lambda,\upsilon\right]\right\},
\end{align*}
where $\lambda$ and $\upsilon$ denote the lower and upper decision threshold, respectively. A decision is then made at each node $k$ according to
\begin{align*}
\text{if } S_k(T_k)\leq\lambda: &\ \text{decide for } \mathcal{H}_0,\\	
\text{if } S_k(T_k)\geq\lambda: &\ \text{decide for } \mathcal{H}_1.
\end{align*}
Since $S_k(T)$ is well-defined, the probability of false alarm can be written as \cite{sahu2016distributed}
\begin{align}
\begin{aligned}
	P_\text{FA} &= P_0(S_k(T) \geq \upsilon)\\
	&\leq \sum_{t=1}^\infty P_0(S_k(t) \geq \upsilon)\\
	&\approx \sum_{t=1}^\infty \mathcal{Q}\left(\frac{\upsilon-\mu_{\eta,0}t}{\sigma_{\eta,0} \sqrt{\xi t}}\right),
	\label{eq:pfa_sk}
	\end{aligned}
\end{align}
where $\mathcal{Q}(x)$ denotes the tail probability of the standard normal distribution. Inequality~\eqref{eq:pfa_sk} holds true as long as the test statistic follows a Gaussian distribution. This is always the case in a shift-in-mean setup since the log-likelihood ratio is also Gaussian distributed. For shift-in-variance tests, the log-likelihood ratio follows a chi-squared distribution with 1 degree of freedom as shown in Fig.~\ref{fig:vartest}(a). Hence, we can use the central limit theorem to state that Eq.~\eqref{eq:pfa_sk} is approximately true after just a few time steps as depicted in Fig.~\ref{fig:vartest}(b). For more details on this, see Section~\ref{sec:pdf_llr}.

Using the property $\mathcal{Q}(x) \leq \frac{1}{2}e^{-\frac{x^2}{2}}$ and taking a similar approach as the authors in \cite{sahu2016distributed}, we obtain
\begin{align*}
\begin{aligned}
	P_\text{FA} &\leq\frac{2e^{\frac{\upsilon\mu_{\eta,0}}{4\sigma_{\eta,0}^2\xi}}}{1-e^{-\frac{\mu_{\eta,0}^2}{2\sigma_{\eta,0}^2\xi}}}.
\end{aligned}	
\end{align*}
Requiring $P_\text{FA} \leq \alpha$ and solving for $\upsilon$ yields the upper threshold
\begin{align}
	\upsilon &\geq \frac{4\sigma_{\eta,0}^2\xi}{\mu_{\eta,0}} \left[ \log\!\left(\frac{\alpha}{2}\right) + \log\!\left( 1-e^{-\frac{\mu_{\eta,0}^2}{2\sigma_{\eta,0}^2\xi}}\right) \right].
	\label{eq:gamma_u}
\end{align}
Repeating the same procedure for the probability of misdetection and requiring $P_\text{MD}\leq \beta$ yields the lower threshold
\begin{align}
	\lambda &\leq \frac{4\sigma_{\eta,1}^2\xi}{\mu_{\eta,1}} \left[ \log\!\left(\frac{\beta}{2}\right) + \log\!\left( 1-e^{-\frac{\mu_{\eta,1}^2}{2\sigma_{\eta,1}^2\xi}}\right) \right].
	\label{eq:gamma_l}
\end{align}
	As mentioned in \cite{sahu2016distributed}, tighter thresholds can be obtained by numerically solving
	\begin{align}
	\begin{aligned}
	\frac{1}{2}\sum_{t=1}^\infty e^{\frac{-\lambda^2 - \mu_{\eta,1}^2t^2 + 2\lambda\mu_{\eta,1}t}{2\sigma_{\eta,1}^2\xi t}} &= \beta\\
	\frac{1}{2}\sum_{t=1}^\infty e^{\frac{-\upsilon^2 - \mu_{\eta,0}^2t^2 + 2\upsilon\mu_{\eta,0}t}{2\sigma_{\eta,0}^2\xi t}}&= \alpha.
	\end{aligned}
\end{align}
The complete derivation is given in Appendix \ref{app:gen_th}.

\section{Robust Distributed Sequential Detection using the LFD-$\boldsymbol{\mathcal{CI}}SPRT$}
\label{sec:lfd-cisprt}
In this section we use to the concept of least-favorable densities (LFDs) to modify the $\mathcal{CI}$SPRT such that it can deal with composite hypotheses arising from distributional uncertainties.

\subsection{Least-favorable Densities (LFDs)}
The set of possible distributions $\mathcal{P}_i$ under hypothesis $\mathcal{H}_i$ can be characterized with the help of Kassam's band model \cite{kassam1981robust,fauss2015old-bands} as
\begin{align*}
	\mathcal{P}_i = \left\{P_i\ \Big|\ p_i^{\prime}\leq p_i\leq p_i^{\prime\prime}\right\},
\end{align*}
i.e., the true density $p_i$ is assumed to lie within a band specified by $p_i^{\prime}$ and $p_i^{\prime\prime}$. A pair of densities $(q_0,q_1)$ within the respective bands is said to be least favorable if they characterize the worst case of a centralized fixed-sample-size test between $\mathcal{H}_0$ and $\mathcal{H}_1$. Using the algorithm in \cite[Table 1]{fauss2015old-bands}, we can iteratively calculate the LFDs as
\begin{align}
\begin{aligned}
	q_0 &= \min\left\{p_0^{\prime\prime}, \max\left\{c_0(\nu q_0 + q_1), p_0^{\prime}\right\}\right\}\\
	q_1 &= \min\left\{p_1^{\prime\prime}, \max\left\{c_1(q_0 + \nu q_1), p_1^{\prime}\right\}\right\}
	\end{aligned}
	\label{eq:lfds}
\end{align}
for some $\nu \geq 0$ and some $c_0, c_1 \in (0,\frac{1}{\nu}]$. We assume uncertainties of the $\varepsilon$-contamination type \cite{zoubir2012robust}, i.e., 
\begin{align}
\begin{aligned}
	p_i &= (1-\varepsilon) p_i^0 + \varepsilon h_i\\
	&= p_i^{\prime} + \varepsilon h_i,
		\label{eq:pdf_cont}
	\end{aligned}
\end{align}
with contamination factor $\varepsilon$, and $p_i^0$ and $h_i$ denoting the nominal and the contamination distribution under $\mathcal{H}_i$, respectively. To represent $\varepsilon$-contamination with Kassam's band model we set $p_0^{\prime\prime} = p_1^{\prime\prime} = \infty$  and $\nu = 0$ in Eq. \eqref{eq:lfds}, which reduces to \cite{fauss2015old-bands}
\begin{align}
\begin{aligned}
	q_0 &= \max\left\{c_0 q_1, p_0^{\prime}\right\}\\
	q_1 &= \max\left\{c_1 q_0, p_1^{\prime}\right\}.
	\end{aligned}
	\label{eq:lfds_eps}
\end{align}
The resulting densities correspond to the LFDs of Huber's clipped likelihood ratio test \cite{huber1965robust,huber1981robust}, which censors outliers and, thus, prevents them from having an unbounded effect on the test. Due to this property, it makes sense to use the centralized, fixed-sample-size LFDs also in the context of distributed sequential detection. While they are not minimax optimal in this case, they induce robustness by limiting the influence of large values at the cost of an increased average run length as we will see in Section~\ref{sec:simulations}.

\subsection{The Robust Test Static and Its Density}
\begin{figure}[tb]
\psfrag{x}[tc][tc][1]{Clipped log-likelihood ratio $\eta_k^\text{c}(t)$}
\psfrag{y}[bc][bc][1]{empirical probability density}
\psfrag{0}[cc][cc][1]{$0$}
\psfrag{0.5}[cc][cc][1]{$0.5$}
\psfrag{-0.5}[cc][cc][1]{$-0.5$}
\psfrag{1}[cc][cc][1]{$1$}
\psfrag{0.2}[cc][cc][1]{$0.2$}
\psfrag{0.4}[cc][cc][1]{$0.4$}
\psfrag{0.6}[cc][cc][1]{$0.6$}
\psfrag{0.8}[cc][cc][1]{$0.8$}
\includegraphics[width=0.49\textwidth]{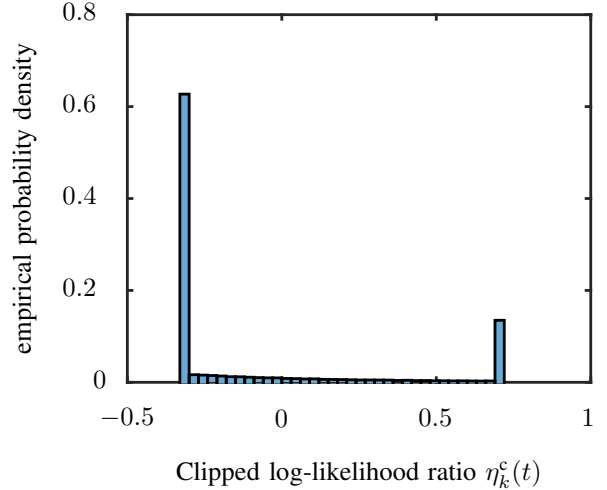}
\caption{Typical histogram of the clipped log-likelihood ratio $\eta_k^\text{c}(t)$ under $\mathcal{H}_0$ and $\varepsilon$-contaminated noise with $\varepsilon=0.1$}
\label{fig:hist_approx}	
\end{figure}

In order to design a robust version of the $\mathcal{CI}$SPRT we replace the log-likelihood ratio $\eta_k(t)$ of agent $k$ at time instant $t$ in Eq. (\ref{eq:sk}) by the corresponding clipped log-likelihood ratio $\eta_k^\text{c}(t)$ with
\begin{align}
	\eta_k^\text{c}(t) = \log\!\left(\frac{q_1(y_k(t))}{q_0(y_k(t))}\right).
\end{align}
This yields a robust test statistic $\check{S}_k(t)$ as
\begin{align}
	\check{S}_k(t) &= \sum_{l\in \mathcal{N}_k\cup\{k\}}w_{kl}\left(\check{S}_l(t-1)+ \eta_l^\text{c}(t)\right).
	\label{eq:sk_rob}
\end{align}
The probability density of $\eta_k^\text{c}(t)$ is shown in the exemplary histogram in Fig.\;\ref{fig:hist_approx}. Considering $\varepsilon$-contamination as defined in Eq.~\eqref{eq:pdf_cont}, the density is composed of two terms. The first one corresponds to the density of the regular log-likelihood ratio under the nominal distribution scaled by $(1-\varepsilon)$. The second one is the probability of drawing an outlier---denoted by $\epsilon$. In the worst case, which is represented by the LFDs, the probability of drawing an outlier is placed at the maximum (minimum) of the log-likelihood ratio under $\mathcal{H}_0  $ ($\mathcal{H}_1$). The probability density is then clipped at $C_0~=~-\log\left(c_0\right)$ and $C_1~=~\log\left(c_1\right)$ to avoid an unbounded influence of outliers. The excess probability that accumulates at the clipping points can be calculated as
\begin{align*}
	&\begin{aligned}
		A_{0,i} &= Q_i(\eta_k(t) \leq C_0)\\
		&=(1-\varepsilon)P^{0}_i(\eta_k(t) \leq C_0)+i\varepsilon\\
			&\approx (1-\varepsilon)\mathcal{Q}\!\left(-\frac{C_0 - \mu_{p^{0}_i}}{\sigma_{p^{0}_i}^2}\right)+i\varepsilon\\
	\end{aligned}
	\\
	&\begin{aligned}
		A_{1,i} &= Q_i(\eta_k(t) \geq C_1)\\
		&= (1-\varepsilon)P^{0}_i(\eta_k(t) \geq C_1)+(1-i)\varepsilon\\
			&\approx (1-\varepsilon)\mathcal{Q}\!\left(\frac{C_1 - \mu_{p^{0}_i}}{\sigma_{p^{0}_i}^2}\right)+(1-i)\varepsilon,		
	\end{aligned}
\end{align*}
where $\mu_{p^{0}_i}$ and $\sigma_{p^{0}_i}^2$ denote the mean and the variance of the nominal distribution under $\mathcal{H}_i, i\in[0,1]$.
\begin{figure*}[!t]
\psfrag{x1}[tc][tc][0.9]{$t=1$}
\psfrag{x2}[tc][tc][0.9]{$t=2$}
\psfrag{x3}[tc][tc][0.9]{$t=3$}
\psfrag{y1}[bc][bc][0.9]{probability density}
\psfrag{y2}[bc][bc][0.9]{}
\psfrag{y3}[bc][bc][0.9]{}
\psfrag{-1}[tc][tc][0.9]{$-1$}
\psfrag{0}[tc][tc][0.9]{$0$}
\psfrag{1}[tc][tc][0.9]{$1$}
\psfrag{2}[tc][tc][0.9]{$2$}
\psfrag{3}[tc][tc][0.9]{$3$}
\psfrag{4}[tc][tc][0.9]{$4$}
\psfrag{5}[tc][tc][0.9]{$5$}
\psfrag{-5}[tc][tc][0.9]{-$5$}
\psfrag{-3}[tc][tc][0.9]{-$3$}
\psfrag{6}[tc][tc][0.9]{$6$}
\psfrag{8}[tc][tc][0.9]{$8$}
\psfrag{10}[tc][tc][0.9]{$10$}
\psfrag{20}[tc][tc][0.9]{$20$}
\psfrag{-20}[tc][tc][0.9]{$-20$}
\psfrag{40}[tc][tc][0.9]{$40$}
\psfrag{0.05}[tc][tc][0.9]{$0.05$}
\psfrag{0.1}[tc][tc][0.9]{$0.1$}
\psfrag{0.15}[tc][tc][0.9]{$0.15$}
\psfrag{0.2}[tc][tc][0.9]{$0.2$}
\includegraphics[width=\textwidth]{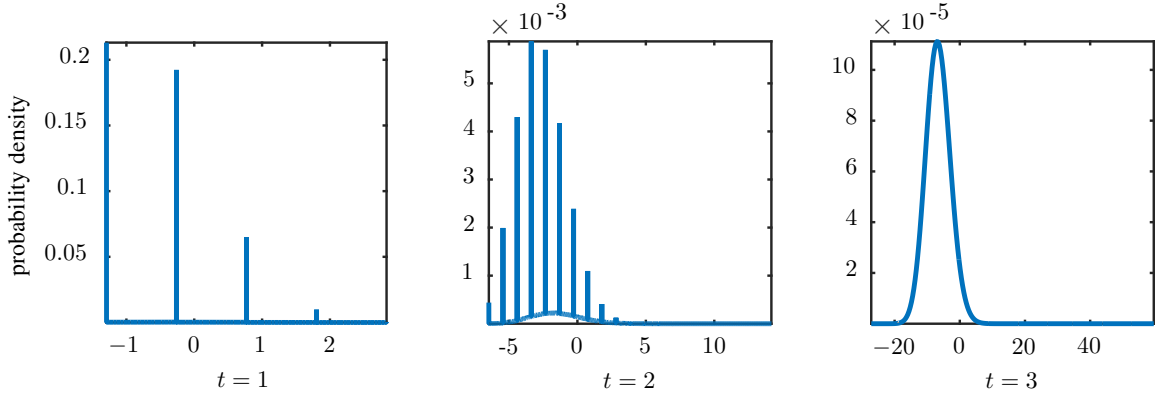}
\caption{Evolution of the probability density function of the robust test statistic $\check{S}_k(t)$ of an agent with 3 neighbors over time}\label{fig:approx_normal}	
\end{figure*}
\subsection{Mean and Variance of the Robust Test Statistic}
The mean and the variance of the robust test statistic can be calculated by finding expressions for the mean $\mu_{\eta^\text{c},i}$ and the variance $\sigma_{\eta^\text{c},i}^2$ of the clipped log-likelihood ratio under $\mathcal{H}_i$ first. Since the distribution is equal for all agents, the superscript $k$ is dropped in the following derivation. We approximate the probability density from Fig.~\ref{fig:hist_approx} by two weighted Kronecker deltas at $C_0$ and $C_1$ with a weighted uniform distribution in between. Note that the uniform distribution is just one convenient possibility to approach the density in this interval of extremely low weight compared to the two Kronecker deltas at the end points. The mean $\mu_{\eta^\text{c},i}$ and the variance $\sigma_{\eta^\text{c},i}^2$ of the robust log-likelihood ratio are calculated according to
\begin{align}
&\begin{aligned}
	\mu_{\eta^\text{c},i} &= E_i\{\eta^\text{c}\} \\&= \int_{\Omega}p_{\eta^\text{c},i}(x) x\;dx\\
	&= \int_{C_0}^{C_1} \left(A_{0,i}\delta(x-C_0) + A_{1,i}\delta(x-C_1)+A_{2,i}\right)x\;dx\\
	&= A_{0,i}C_0 + A_{1,i}C_1 + \frac{1}{2} \left[A_{2,i}x^2\right]_{C_0}^{C_1}\\
	&= A_{0,i}C_0 + A_{1,i}C_1+A_{2,i}\frac{C_1^2-C_0^2}{2},
	\end{aligned}
	\label{eq:mu_eta_clipped}
\end{align}
and
\begin{align}
	&\begin{aligned}
	\sigma_{\eta^\text{c},i}^2 &= E_i\{\left(\eta^\text{c}\right)^2\} - \mu_{\eta^\text{c},i}^2 \\&= \int_{\Omega}p_{\eta^\text{c},i}(x) x^2\;dx - \mu_{\eta^\text{c},i}^2\\
	&= \int_{C_0}^{C_1} \left(A_{0,i}\delta(x-C_0) + A_{1,i}\delta(x-C_1)+A_{2,i}\right)x^2\;dx \\&\qquad- \mu_{\eta^\text{c},i}^2\\
	&= A_{0,i}C_0^2 + A_{1,i}C_1^2 +A_{2,i}\frac{C_1^3-C_0^3}{3} - \mu_{\eta^\text{c},i}^2,
	\end{aligned}
	\label{eq:var_eta_clipped}
\end{align}
respectively, with $A_{2,i}~=\frac{(1-A_{0,i}-A_{1,i})}{C_1-C_0}$.

The derivation of the mean and the variance of the non-robust test statistic $S_k(t)$ in Section \ref{sec:mu_var_sk} is based on the assumption of a Gaussian-distributed log-likelihood ratio $\eta_k(t)$. While this assumption does not hold for the clipped log-likelihood ratio $\eta_k^\text{c}(t)$, we can resort to the central limit theorem to state that the robust test statistic $\check{S}_k(t)$ is approximately normal \cite{lehmann2005testing,kay2006intuitive}. The evolution of the probability density function of $\check{S}_k(t)$ over time is depicted in Fig.~\ref{fig:approx_normal} for an agent with three neighbors. As can be seen, the data exchange over the neighborhood causes the probability density function to become approximately Gaussian already after the first few time instants. An even faster convergence can be observed in denser networks. Hence, the mean and variance of $\check{S}_k(t)$ can be calculated by replacing $\mu_{\eta,i}$ and $\sigma_{\eta,i}^2$ in (\ref{eq:mu_sk}) and (\ref{eq:var_sk}) with their robust counterparts, i.e.,
\begin{align}
	E_i\{\check{S}_k(t)\} &= \mu_{\eta^\text{c},i} t\label{eq:mean_s_rob}\\
	\text{Var}_i\{\check{S}_k(t)\} &\leq \sigma^2_{\eta^\text{c},i} \xi t.\label{eq:var_s_rob}
\end{align}

\subsection{Robust Decision Thresholds}
\label{sec:rob_dec_th}
The mean and the variance of the robust test statistic $\check{S}_k(t)$ in Eqs. \eqref{eq:mean_s_rob} and \eqref{eq:var_s_rob} have the same form as those of the non-robust test statistic $S_k(t)$ in Eqs. \eqref{eq:mu_sk} and \eqref{eq:var_sk}. Therefore, we can derive robust decision thresholds by following the paradigm from Section \ref{sec:deriv_th}, i.e., by replacing the mean and the variance of the log-likelihood ratio $\eta_k(t)$ in Eqs. \eqref{eq:gamma_u} and \eqref{eq:gamma_l} with those of the clipped log-likelihood ratio $\eta_k^\text{c}(t)$ from Eqs. \eqref{eq:mu_eta_clipped} and \eqref{eq:var_eta_clipped}. Thus, we obtain
\begin{align}
\begin{aligned}
	\check{\upsilon} &\geq \frac{4\sigma_{\eta^\text{c},0}^2\xi}{\mu_{\eta^\text{c},0}} \left[ \log\!\left(\frac{\alpha}{2}\right) + \log\!\left( 1-e^{-\frac{\mu_{\eta^\text{c},0}^2}{2\sigma_{\eta^\text{c},0}^2\xi}}\right) \right]\\
	\check{\lambda} &\leq \frac{4\sigma_{\eta^\text{c},1}^2\xi}{\mu_{\eta^\text{c},1}} \left[ \log\!\left(\frac{\beta}{2}\right) + \log\!\left( 1-e^{-\frac{\mu_{\eta^\text{c},1}^2}{2\sigma_{\eta^\text{c},1}^2\xi}}\right) \right],
\end{aligned}
	\label{eq:gamma_robust}
\end{align}
with tighter bounds arising from numerically evaluating 
\begin{align}
	\begin{aligned}
	\frac{1}{2}\sum_{t=1}^\infty e^{\frac{-\lambda^2 - \mu_{\eta^\text{c},1}^2t^2 + 2\lambda\mu_{\eta^\text{c},1}t}{2\sigma_{\eta^\text{c},1}^2\xi t}} &= \beta\\
	\frac{1}{2}\sum_{t=1}^\infty e^{\frac{-\upsilon^2 - \mu_{\eta^\text{c},0}^2t^2 + 2\upsilon\mu_{\eta^\text{c},0}t}{2\sigma_{\eta^\text{c},0}^2\xi t}}&= \alpha.
	\end{aligned}
\end{align}

%

\section{Robust Distributed Sequential Detection using Robust Estimators}
\label{sec:rob_est}
In this section we show how to leverage the diversity of a distributed sensor network along with robust estimators to introduce robustness through the update equation of the $\mathcal{CI}$SPRT. We start by reformulating Eq.~\eqref{eq:sk} as
\begin{align}
	S_k(t) &= \sum_{l\in \mathcal{N}_k\cup\{k\}}w_{kl}\left(S_l(t-1)\right)+ \hat{\eta}_k(t),
	\label{eq:sk_rest}
\end{align}
	with $\hat{\eta}_k(t)$ denoting the weighted average of the collective innovations terms of node $k$ and its neighborhood at time $t$. When no a priori knowledge about the reliability of the nodes is available, a common choice is to weight all the information equally. This leads to $\hat{\eta}_k(t)$ being the sample mean
	\begin{align}
		\hat{\eta}_k^\text{mean}(t) = \frac{1}{|\mathcal{N}_k\cup\{k\}|}\sum_{l\in\mathcal{N}_k\cup\{k\}}\eta_l(t),
	\end{align}
	which is a non-robust estimator \cite{zoubir2012robust}. Since the update equation is recursive, replacing the sample mean with a robust alternative will robustify the consensus part as well and, thus, yield a test statistic that can handle distributional uncertainties. An advantage of introducing robustness in this manner instead of using LFDs as detailed in the previous section is the fact that the censoring takes place one stage later. Instead of clipping the log-likelihood ratio directly, the effect of large values on the innovations term is bounded by using a robust estimator in the combination rule. Thus, the thresholds and decision rules of the original $\mathcal{CI}$SPRT, which are based on the mean and the variance of the log-likelihood ratio, remain valid. 
	
	A first attempt at using this approach was presented in \cite{hou2017robust}, where we successfully used the median, the M-estimator, and the sample myriad \cite{gonzalez1996weighted, gonzalez2002statistically-efficient} for sequential detection. In the sequel, we will briefly summarize these algorithms and investigate their suitability for different binary hypothesis tests.
		
	\subsection{The Median-$\mathcal{CI}$SPRT}
	A straightforward way of replacing the sample mean in Eq. \eqref{eq:sk_rest} with a robust alternative is to use the median $\hat{\eta}_k^\text{median}(t)$. The estimate of the innovations term is calculated as 
	\begin{align}
	\hat{\eta}_k^\text{median}(t) = \begin{cases}
									\boldsymbol{\eta}_k(\frac{|\mathcal{N}_k|+1}{2})&\!,|\mathcal{N}_k|\ \text{even}\\
									\frac{1}{2}\left(\boldsymbol{\eta}_k(\frac{|\mathcal{N}_k|+1}{2})+\boldsymbol{\eta}_k(\frac{|\mathcal{N}_k|+1}{2}+1)\right)&\!,|\mathcal{N}_k|\  \text{odd}
 \end{cases},
 \label{eq:eta_median}
	\end{align}
	with $\boldsymbol{\eta}_k(t)$ denoting the vector of the log-likelihood ratios of node $k$ and its neighbors sorted in ascending order.
	\subsection{The M-$\mathcal{CI}$SPRT}
	The M-$\mathcal{CI}$SPRT is obtained by using an M-estimate of the neighborhood-wide innovations part in Equation \eqref{eq:sk_rest}. Intuitively speaking, the M-estimator provides a weighted average with weights given by \cite{zoubir2012robust}
	\begin{equation*}
	W(x)=\begin{cases}
	\frac{\psi (x)}{x}&,x\neq 0\\ 
	\psi^{\prime}(0)&,x= 0
	\end{cases},
	\end{equation*}
	where $ \psi (x) $ is a score function and $\psi^{\prime}(x)$ its first derivative. In this work we consider Huber's 
	score function defined as \cite{zoubir2012robust,huber1981robust}
	\begin{align*}
		\psi_\text{Hub}(x) = 
		\begin{cases}
	x &,|x|\leq c_\text{Hub}\\ 
	c_\text{Hub}\text{sign}(x)&,|x|> c_\text{Hub}
	\end{cases},\\
	\end{align*}
	for some postive constant $c_\text{Hub}$.
	
	The M-estimate of the innovations term is obtained by recursively calculating \cite{zoubir2012robust,huber1981robust}
	\begin{align}
		w_{k}(t,i) &= W\left(\frac{\eta_k(t)-\hat{\eta}_k^\text{M}(t,i)}{\hat{\sigma}(\boldsymbol{\eta}_k(t))}\right)\\
		\hat{\eta}_k^\text{M}(t,i+1) &= \frac{\sum_{l\in\mathcal{N}_k\cup\{k\}}w_l(i)\eta_l(t,i)}{\sum_{l\in\mathcal{N}_k\cup\{k\}}w_l(i)}
	\end{align}
	until $\frac{|\hat{\eta}_k^\text{M}(t,i+1)-\hat{\eta}_k^\text{M}(t,i)|}{\hat{\sigma}(\boldsymbol{\eta}_k(t))}~<~\varepsilon$ for a small, positive constant $\varepsilon$. The algorithm is initialized by setting $\hat{\eta}_k^\text{M}(t,0)~=~\hat{\eta}_k^\text{median}(t)$ and estimating the scale using the normalized median standard deviation according to \cite{zoubir2012robust}
	\begin{align*}
		\hat{\sigma}_\text{mad}(\boldsymbol{\eta}_k(t)) = 1.483 \cdot \text{median}\left(|\boldsymbol{\eta}_k(t)-\hat{\eta}_k^\text{median}(t)|\right),
	\end{align*}
	where the median is calculated as in Eq. \eqref{eq:eta_median}.

	\subsection{The Myriad-$\mathcal{CI}$SPRT}
	The third robust estimator we consider in this work is the sample myriad, which estimates the innovations term according to \cite{gonzalez1996weighted,gonzalez2002statistically-efficient}
	 \begin{equation}
	    \hat{\eta}_k^\text{myriad}(t) =\mathrm{arg}\min_{\eta }\prod _{{l\in\mathcal{N}_k\cup\{k\}}} \left [ m^{2}+\left ( \eta_l(t)-\eta  \right )^{2} \right ]
	 \end{equation}
	where $ m $ is a freely tunable parameter. A common choice is to set $m = \hat{\sigma}_\text{mad}(\boldsymbol{\eta}(t))$ \cite{gonzalez2002statistically-efficient}.
\begin{figure*}[tb]
	\centering
	\psfrag{CISPRT}[l][l][0.6]{$\mathcal{CI}$SPRT}
	\psfrag{LFD-CISPRT}[l][l][0.6]{LFD-$\mathcal{CI}$SPRT}
	\psfrag{Median-CISPRT}[l][l][0.6]{Median-$\mathcal{CI}$SPRT}
	\psfrag{M-CISPRT}[l][l][0.6]{M-$\mathcal{CI}$SPRT}
	\psfrag{Myriad-CISPRT}[l][l][0.6]{Myriad-$\mathcal{CI}$SPRT}
	\psfrag{req. PMD}[l][l][0.6]{req. error prob.}
	\psfrag{x}[c][c][0.9]{Required error probability}
	\psfrag{y}[b][c][0.9]{Average run length}
	\subfloat{
	\includegraphics[width=0.5\textwidth]{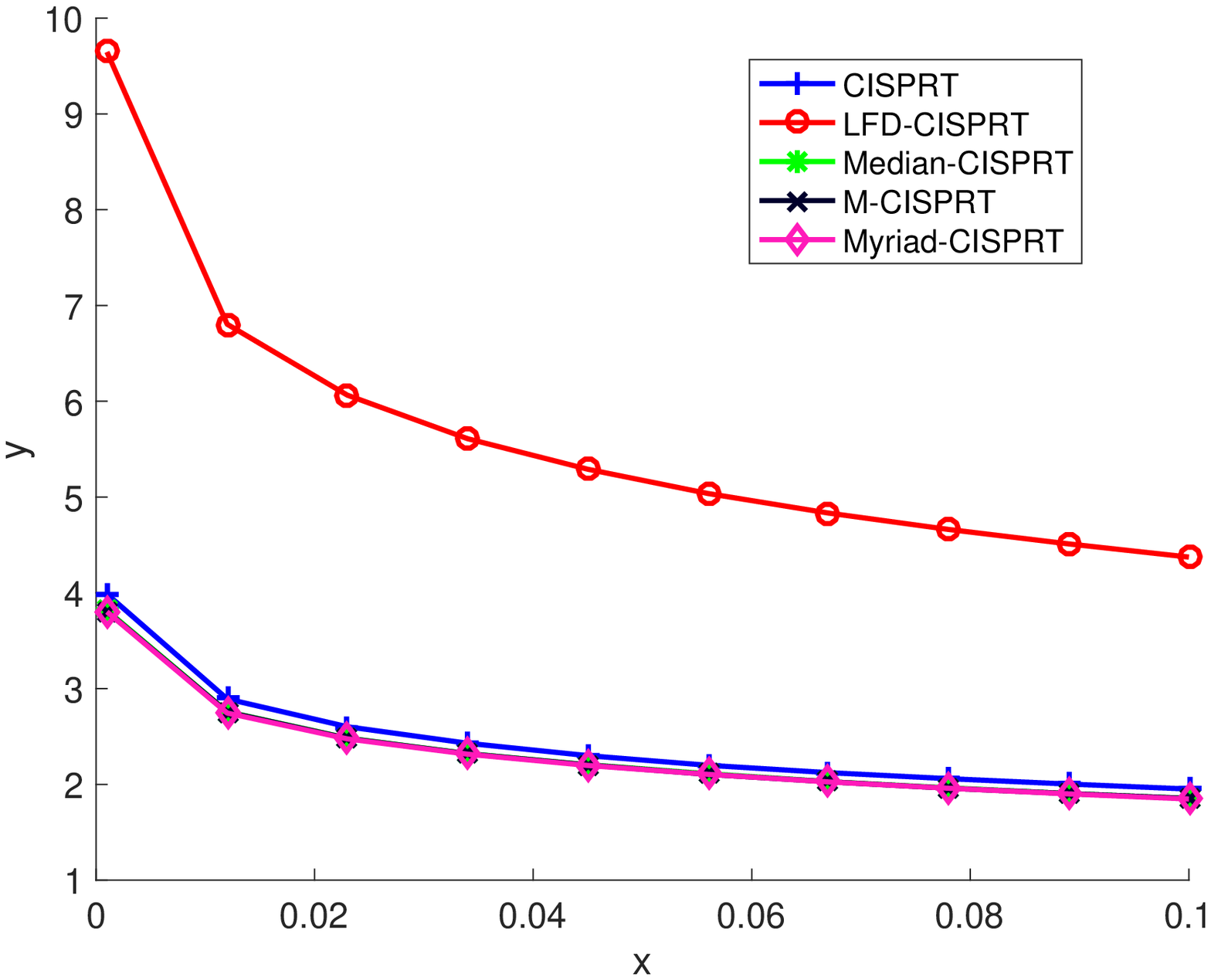}
	}
	\psfrag{y}[b][c][0.9]{Empirical error probability}
	\subfloat{
	\includegraphics[width=0.5\textwidth]{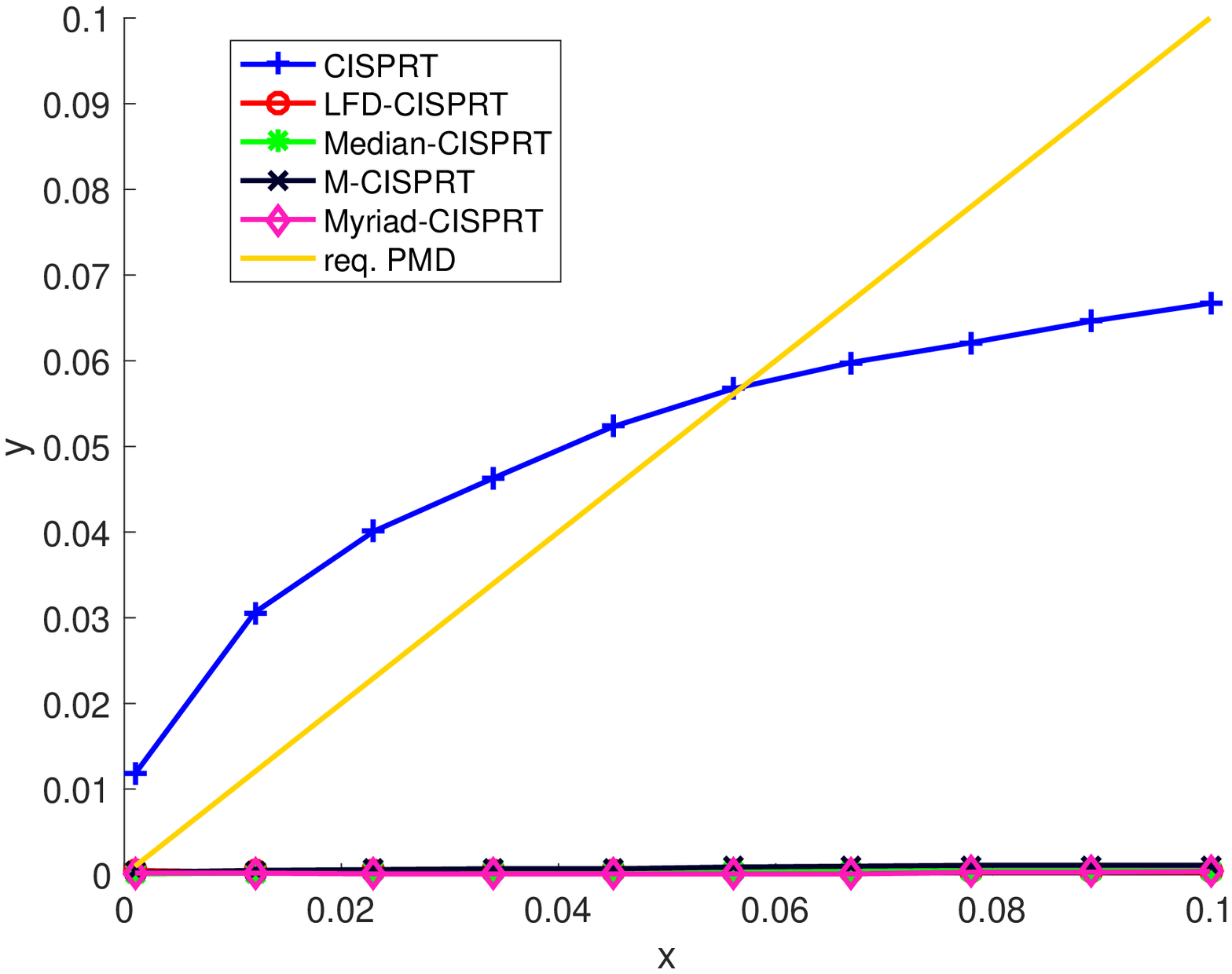}
	}
	\caption{Simulation results for the shift-in-mean test. Due to the symmetry, the results are equal under $\mathcal{H}_0$ and $\mathcal{H}_1$}
	\label{fig:res_mean}
\end{figure*}	
	\subsection{The Probability Density Function of the Log-Likelihood Ratio}
	\label{sec:pdf_llr}
	We are concerned with shift-in-mean as well as shift-in-variance tests. In order to investigate the suitability of the proposed detectors in these two cases, we take a look at the probability density function of the estimator input, i.e., the neighborhood innovations of node $k$. Since uncontaminated measurements are assumed to be Gaussian, we can write
	\begin{align*}
	 y_k(t) = \mu^\circ + x(t)\sigma^\circ,
	 \end{align*}
	 where $x(t)\sim\mathcal{N}(0,1)$, and $\mu^\circ$ and $\sigma^\circ$ denote the true mean and standard deviation of $y_k(t)$. The log-likelihood ratio of node $k$ as defined in Eq.~\eqref{eq:eta_k} now becomes
	\begin{align*}
	\begin{aligned}
		\intertext{\textit{Shift-in-mean} ($\sigma_0=\sigma_1=\sigma$):}
		\eta_k(t) &= \frac{\sigma_1^2\left(y_k(t)-\mu_0\right)^2-\sigma_0^2\left(y_k(t)-\mu_1\right)^2}{2\sigma_0^2\sigma_1^2} + \log\!\left(\frac{\sigma_0}{\sigma_1}\right)\\
		&= \frac{\left(y_k(t)-\mu_0\right)^2-\left(y_k(t)-\mu_1\right)^2}{2\sigma^2}\\
		&= \frac{\left(\sigma^\circ x(t)+\left(\mu^\circ-\mu_0\right)\right)^2-\left(\sigma^\circ x(t)+\left(\mu^\circ-\mu_1\right)\right)^2}{2\sigma^2}\\
		&= x(t)\frac{\sigma^\circ}{\sigma^2}\left(\mu_1-\mu_0\right) + \frac{\left(\mu^\circ-\mu_0\right)^2-\left(\mu^\circ-\mu_1\right)^2}{2\sigma^2}\\
		&= ax(t) + b.
		\end{aligned}\\
		\begin{aligned}
		\intertext{\textit{Shift-in-variance} ($\mu_0=\mu_1=\mu^\circ=0$):}
		\eta_k(t) &= \frac{\sigma_1^2\left(y_k(t)-\mu_0\right)^2-\sigma_0^2\left(y_k(t)-\mu_1\right)^2}{2\sigma_0^2\sigma_1^2} + \log\!\left(\frac{\sigma_0}{\sigma_1}\right)\\
		&= \frac{\sigma_1^2y_k(t)^2-\sigma_0^2y_k(t)^2}{2\sigma_0^2\sigma_1^2} + \log\!\left(\frac{\sigma_0}{\sigma_1}\right)\\
		&= y_k(t)^2\frac{\sigma_1^2-\sigma_0^2}{2\sigma_0^2\sigma_1^2} + \log\!\left(\frac{\sigma_0}{\sigma_1}\right)\\
		&= \left(\sigma^\circ\right)^2x^2(t)\frac{\sigma_1^2-\sigma_0^2}{2\sigma_0^2\sigma_1^2} + \log\!\left(\frac{\sigma_0}{\sigma_1}\right)\\
		&= c x^2(t) + d.
		\end{aligned}
	\end{align*}
	The values of $a, b, c,$ and $d$ are clear from the context. Thus, in the shift-in-mean test $\eta_k(t)$ follows a Gaussian distribution. In the shift-in-variance test, however, this is not the case. Since $x(t)$ follows the standard normal distribution, $x^2(t)$ is chi-squared distributed with one degree of freedom, i.e., $x^2(t)\sim\chi^2_1$. Hence, the log-likelihood ratio follows a scaled and shifted $\chi^2_1$ distribution, which is not symmetric but skewed.
	
	Regarding our proposed algorithms, this has the following implication: Since the median is only a robust estimator for the mean of symmetric distributions, the Median-$\mathcal{CI}$SPRT is not suitable for general shift-in-variance problems. It might give correct detection results for certain parameter choices as can be seen in the promising simulation results from \cite{hou2017robust}, but we cannot guarantee a reliable performance for arbitrary shift-in-variance tests. Therefore, we will consider the Median-$\mathcal{CI}$SPRT only for shift-in-mean tests.
\begin{figure*}[tb]
	\centering
	\psfrag{CISPRT}[l][l][0.6]{$\mathcal{CI}$SPRT}
	\psfrag{LFD-CISPRT}[l][l][0.6]{LFD-$\mathcal{CI}$SPRT}
	\psfrag{Median-CISPRT}[l][l][0.6]{Median-$\mathcal{CI}$SPRT}
	\psfrag{M-CISPRT}[l][l][0.6]{M-$\mathcal{CI}$SPRT}
	\psfrag{Myriad-CISPRT}[l][l][0.6]{Myriad-$\mathcal{CI}$SPRT}
	\psfrag{req. PMD}[l][l][0.6]{req. PMD}
	\psfrag{req. PFA}[l][l][0.6]{req. PFA}
	\psfrag{x}[c][c][0.9]{Required error probability}
	\psfrag{y}[b][c][0.9]{Average run length}
	\subfloat{
	\includegraphics[width=0.5\textwidth]{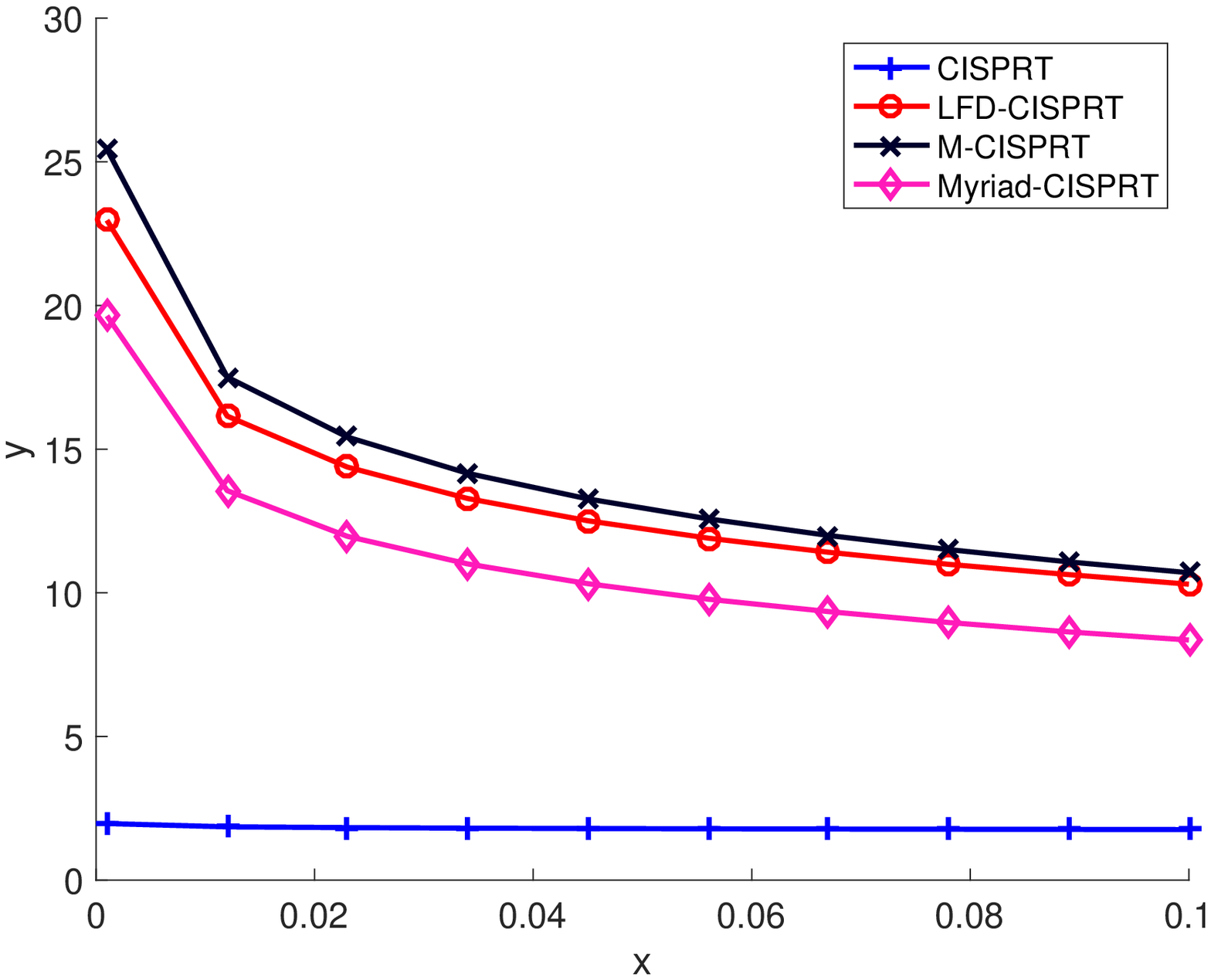}
	}
	\psfrag{x}[c][c][0.9]{Required probability of false alarm}
	\psfrag{y}[b][c][0.9]{Empirical probability of false alarm}
	\subfloat{
	\includegraphics[width=0.5\textwidth]{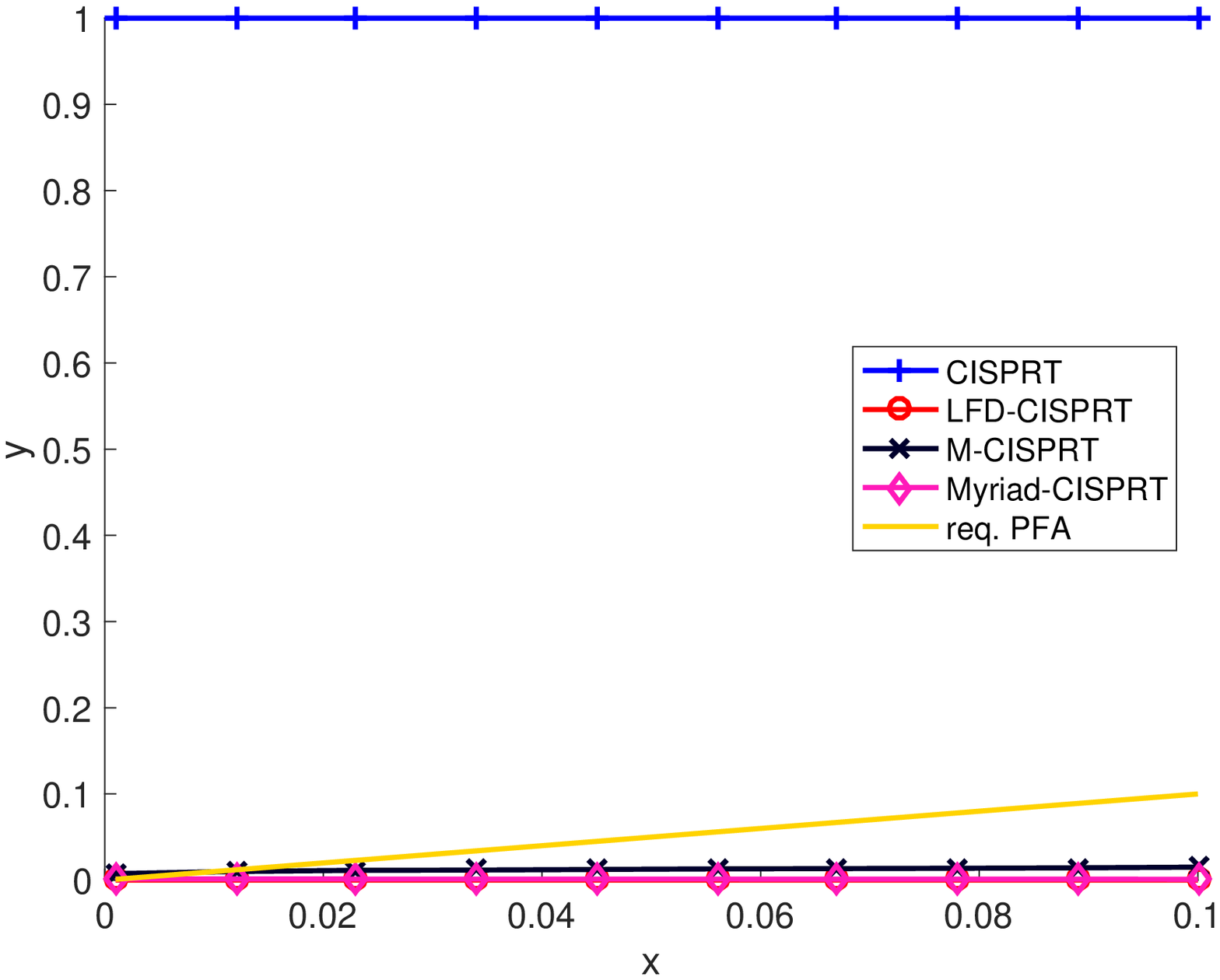}
	}\\
	\psfrag{x}[c][c][0.9]{Required error probability}
	\psfrag{y}[b][c][0.9]{Average run length}
	\subfloat{
	\includegraphics[width=0.5\textwidth]{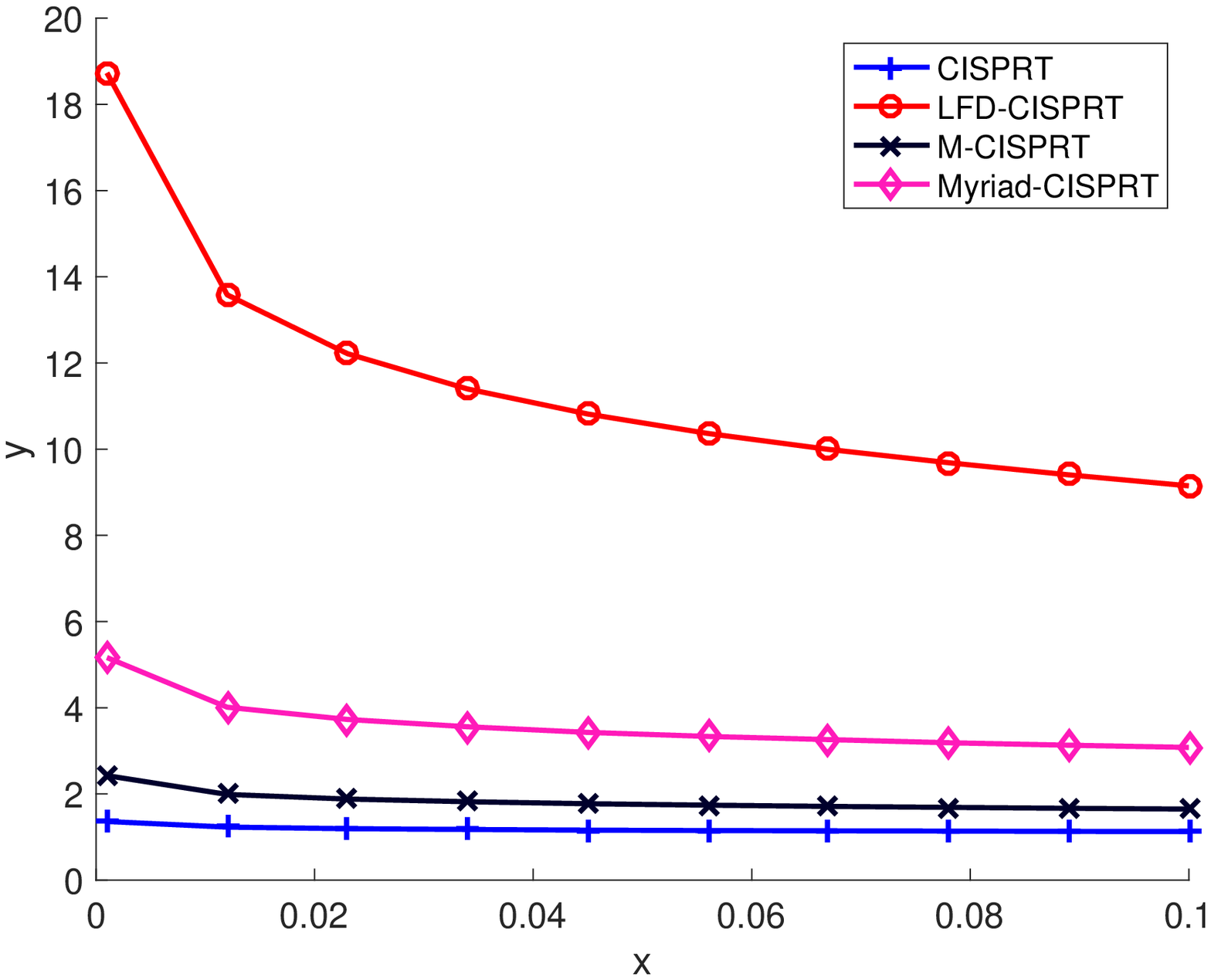}
	}
	\psfrag{x}[c][c][0.9]{Required probability of misdetection}
	\psfrag{y}[b][c][0.9]{Empirical probability of misdetection}
	\subfloat{
	\includegraphics[width=0.5\textwidth]{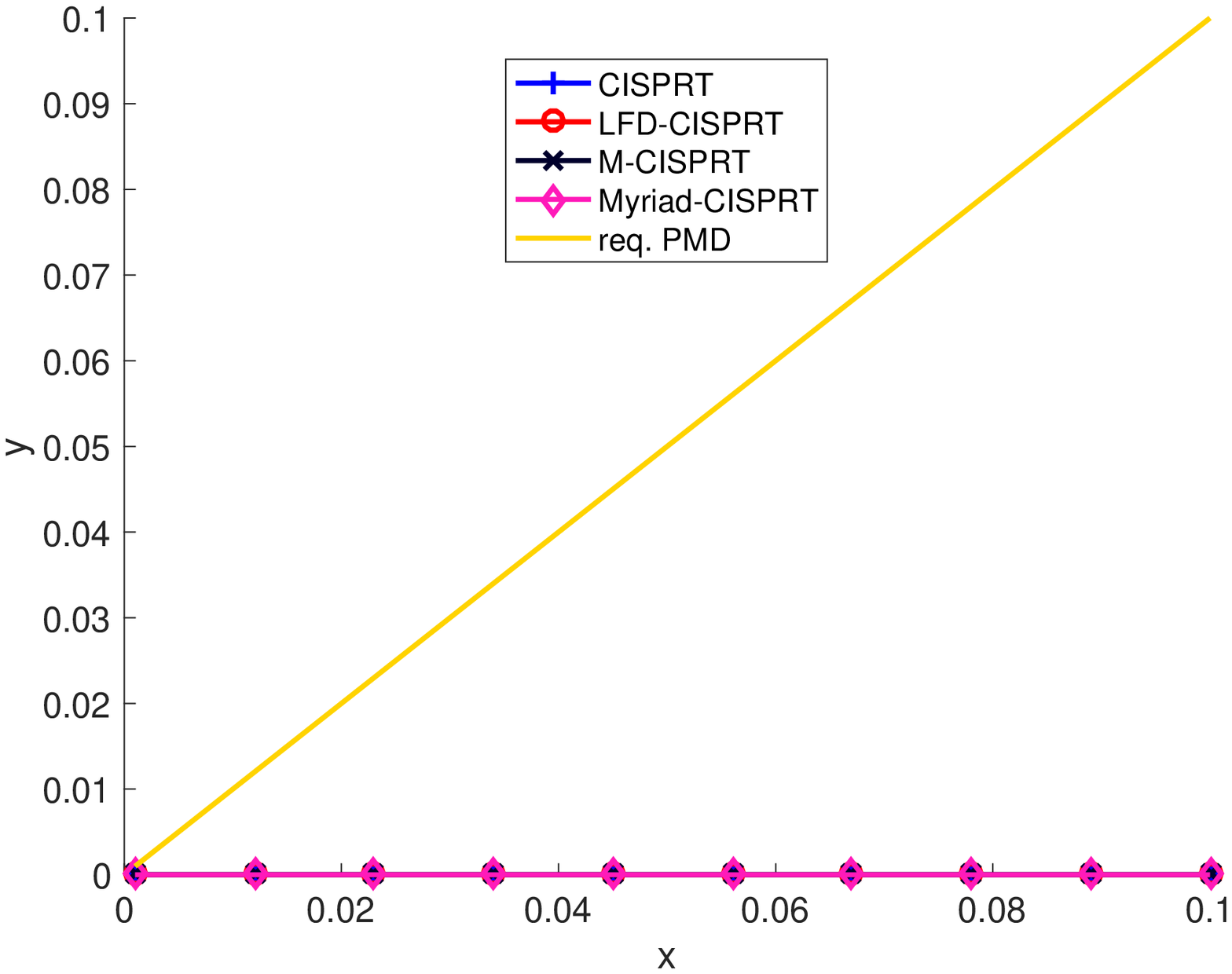}
	}
	\caption{Simulation results for the shift-in-variance test under $\mathcal{H}_0$ (upper row) and $\mathcal{H}_1$ (lower row)}
	\label{fig:res_var}
\end{figure*}	 
\section{Simulations}
\label{sec:simulations}

In this section, we evaluate and compare the performance of the proposed version of the $\mathcal{CI}$SPRT and the proposed robust detectors in the face of $\varepsilon$-contaminated noise. To this end, we consider the two different test scenarios introduced in Section \ref{sec:probform}. In both scenarios, we use a network of $N=20$ agents with uniformly distributed $x$- and $y$-coordinates on the interval $[0,1]$. Agents within a radius of $g = 0.6$ are considered to be neighbors. The required false alarm and misdetection probabilities are assumed to be equal, ranging from $10^{-3}$ to $10^{-1}$. To assess the performance of the different algorithms, we evaluate the average run length as well as the empirical error probabilities, i.e., the probability of false alarm when $\mathcal{H}_0$ is true and the probability of misdetection under $\mathcal{H}_1$. The results are averaged over $N_\text{MC}=10\;000$ Monte Carlo runs.

\subsection{Scenario 1: Shift-in-Mean Test}

In the first scenario, we consider a shift-in-mean problem where the objective is to decide between $\mu_0~=~-1$ and $\mu_1~=~1$ based on measurements that are disturbed by $\varepsilon$-contaminated noise. Here, $\sigma^2 = 2$, the amount of contamination is $\varepsilon = 0.1$, and contaminated measurements suffer from a ten times higher variance.

The simulation results are shown in Fig.~\ref{fig:res_mean}. Due to the symmetry of the problem, the results are equal under both hypotheses. We observe that our proposed robust detection algorithms all meet and even fall below the required error probability while the $\mathcal{CI}$SPRT fails as the requirements get more stringent. At the same time, the Median-$\mathcal{CI}$SPRT, the M-$\mathcal{CI}$SPRT, and the Myriad-$\mathcal{CI}$SPRT exhibit a slightly lower average run length than the $\mathcal{CI}$SPRT. Hence, the robustness property does not come at the cost of a higher testing time. This is in contrast to our results from \cite{hou2017robust}, where the robustification via robust estimators came with a premium in average run length. This effect can be explained by the different decision thresholds due to the different weighting matrices. The decision thresholds in \cite{hou2017robust} are tighter than the ones proposed in this work but they are derived based on certain properties of the weighting matrix that are not meaningful in distributed detection setups as discussed in Section~\ref{sec:w}. Thus, we can conclude that in a common shift-in-mean test, the introduction of robustness through robust estimators does not increase the average run length. The LFD-$\mathcal{CI}$SPRT, in contrast, needs more than twice the testing time than the $\mathcal{CI}$SPRT. This is in line with the results from \cite{leonard2017robust} and \cite{hou2017robust} and due to the fact that the LFDs optimize the algorithm for the worst case, which is not attained by randomly sampling from an $\varepsilon$-contaminated distribution.

As far as the required error probabilities are concerned, all robust algorithms overachieve, i.e., they do not take advantage of the tolerable number of errors but deliver a probability of false alarm and misdetection close to zero. This can be explained by the way the decision thresholds are derived in Sections~\ref{sec:deriv_th} and \ref{sec:rob_dec_th}. As mentioned in \cite{sahu2016distributed}, the approximations required to find a closed-form solution result in thresholds that are sufficient but not optimal.

\subsection{Scenario 2: Shift-in-Variance Test}
The second scenario is a shift-in-variance test. We test for the presence or absence of a signal with variance $\sigma_x^2 = 4$. The noise variance is $\sigma_n^2 = 1$, the amount of contamination is $\varepsilon = 0.1$, and contaminated measurements suffer from a ten times higher variance.

The simulation results are shown in Fig.~\ref{fig:res_var}, where the upper row considers the case where $\mathcal{H}_0$ is true and the lower one pertains to $\mathcal{H}_1$. Under $\mathcal{H}_0$, the $\mathcal{CI}$SPRT breaks down while all robust algorithms meet and even fall below the required error probabilities due to the sufficient conditions on the decision thresholds. Under $\mathcal{H}_1$, however, all algorithms---even the non-robust $\mathcal{CI}$SPRT---meet the error requirements. This is an expected result since, in a shift-in-variance test, outliers, i.e., very large values, actually help in correctly deciding for $\mathcal{H}_1$. As far as the average run length is concerned, the robust algorithms exhibit a five to twelve times larger testing time in the more difficult case where $\mathcal{H}_0$ is true. Moreover, the LFD-$\mathcal{CI}$SPRT is in line with the other robust detectors, which indicates that the considered scenario approaches the worst case. Under $\mathcal{H}_1$, i.e., in the easier case, the average run length of the robust estimator approach is just a few time instants larger than that of the non-robust $\mathcal{CI}$SPRT while the LFD-$\mathcal{CI}$SPRT, again, needs considerably longer to complete the test.

\section{Conclusion}
\label{sec:conclusion}
In this work, we presented a general formulation of the $\mathcal{CI}$SPRT that is not only suitable for sequential binary hypothesis tests but also considers a network structure that is meaningful in the context of distributed detection. Furthermore, we presented two different approaches for robustifying the $\mathcal{CI}$SPRT and proposed four different robust sequential detection algorithms, namely, the LFD-$\mathcal{CI}$SPRT, the Median-$\mathcal{CI}$SPRT, the M-$\mathcal{CI}$SPRT, and the Myriad-$\mathcal{CI}$SPRT. After investigating their suitability for different test setups, we verified, evaluated, and compared their performance in a shift-in-mean and a shift-in-variance test. Our simulation results showed that the proposed detectors are robust against outliers of the $\varepsilon$-contamination type at no or minimal extra cost in terms of the average run length. Only the LFD-$\mathcal{CI}$SPRT comes with a notable increase in testing time due to its focus on the worst-case.

\clearpage
\onecolumn
\appendices

\section{Mean and Variance of the Log-likelihood Ratio}
\label{app:mu_var_eta}

In the following, we derive the mean and the variance of the log-likelihood ratio under the null hypothesis. The derivation under the alternative is analogous. To this end, we make use of the identities
\begin{align*}
	E\{y^2\} &= \mu_y^2 + \sigma_y^2\qquad
	E\{y^3\} = \mu_y^3 + 3\mu_y\sigma_y^2\qquad
	E\{y^4\} = \mu_y^4 + 6\mu_y^2\sigma_y^2 + 3\sigma_y^4.
\end{align*}
Note that we drop the superscript $k$ since the measurements at each agent are assumed to be independently and identically distributed. For the sake of simplicity, the time dependence is omitted as well. 
\begin{align*}
	\mu_{\eta,0} &= E_0\left\{\frac{\sigma_1^2\left(y-\mu_0\right)^2-\sigma_0^2\left(y-\mu_1\right)^2}{2\sigma_0^2\sigma_1^2} + \log\!\left(\frac{\sigma_0}{\sigma_1}\right)\right\}\\
	&= E_0\left\{\frac{\sigma_1^2\left(y^2-2\mu_0y+\mu_0^2\right)-\sigma_0^2\left(y^2-2\mu_1y+\mu_1^2\right)}{2\sigma_0^2\sigma_1^2} + \log\!\left(\frac{\sigma_0}{\sigma_1}\right)\right\}\\
	&= \frac{\sigma_1^2\left(E_0\{y^2\}-2\mu_0E_0\{y\}+\mu_0^2\right)}{2\sigma_0^2\sigma_1^2}-\frac{\sigma_0^2\left(E_0\{y^2\}-2\mu_1E_0\{y\}+\mu_1^2\right)}{2\sigma_0^2\sigma_1^2} + \log\!\left(\frac{\sigma_0}{\sigma_1}\right)\\
	&= \frac{\sigma_1^2\left(\mu_0^2 + \sigma_0^2-2\mu_0^2+\mu_0^2\right)}{2\sigma_0^2\sigma_1^2}-\frac{\sigma_0^2\left(\mu_0^2+\sigma_0^2-2\mu_0\mu_1+\mu_1^2\right)}{2\sigma_0^2\sigma_1^2} + \log\!\left(\frac{\sigma_0}{\sigma_1}\right)\\
	&= \frac{\sigma_1^2\sigma_0^2-\sigma_0^2\left(\mu_0^2+\sigma_0^2-2\mu_0\mu_1+\mu_1^2\right)}{2\sigma_0^2\sigma_1^2}+ \log\!\left(\frac{\sigma_0}{\sigma_1}\right)\\
	&= \underbrace{-\frac{\mu_0^2+\mu_1^2-2\mu_0\mu_1+\sigma_0^2-\sigma_1^2}{2\sigma_1^2}}_{Z_1}+ \underbrace{\log\!\left(\frac{\sigma_0}{\sigma_1}\right)}_{Z_2}
	\end{align*}
\begin{align*}
	E_0\{\eta^2\} &= E_0\left\{\left(\frac{\sigma_1^2\left(y-\mu_0\right)^2-\sigma_0^2\left(y-\mu_1\right)^2}{2\sigma_0^2\sigma_1^2} + \log\!\left(\frac{\sigma_0}{\sigma_1}\right)\right)^2\right\}\\
	&= E_0\left\{\frac{\left(\sigma_1^2\left(y-\mu_0\right)^2-\sigma_0^2\left(y-\mu_1\right)^2\right)^2}{4\sigma_0^4\sigma_1^4} + \log\!\left(\frac{\sigma_0}{\sigma_1}\right)^2 + 2\frac{\sigma_1^2\left(y-\mu_0\right)^2-\sigma_0^2\left(y-\mu_1\right)^2}{2\sigma_0^2\sigma_1^2}\log\!\left(\frac{\sigma_0}{\sigma_1}\right)\right\}\\
	&= \overbrace{E_0\left\{\frac{\left(\sigma_1^2\left(y-\mu_0\right)^2-\sigma_0^2\left(y-\mu_1\right)^2\right)^2}{4\sigma_0^4\sigma_1^4}\right\}}^{Z_3} + \log\!\left(\frac{\sigma_0}{\sigma_1}\right)^2 + 2E_0\left\{\frac{\sigma_1^2\left(y-\mu_0\right)^2-\sigma_0^2\left(y-\mu_1\right)^2}{2\sigma_0^2\sigma_1^2}\right\}\log\!\left(\frac{\sigma_0}{\sigma_1}\right)\\
	&= Z_3 + Z_2^2 + 2Z_1Z_2
\end{align*}
\begin{align*}
	Z_3 &= E_0\left\{\frac{\left(\sigma_1^2\left(y-\mu_0\right)^2-\sigma_0^2\left(y-\mu_1\right)^2\right)^2}{4\sigma_0^4\sigma_1^4}\right\}\\
	&= E_0\left\{\frac{\left(\sigma_1^2\left(y^2-2\mu_0y+\mu_0^2\right)-\sigma_0^2\left(y^2-2\mu_1y+\mu_1^2\right)\right)^2}{4\sigma_0^4\sigma_1^4}\right\}\\
	&= E_0\left\{\frac{\sigma_1^4\left(y^2-2\mu_0y+\mu_0^2\right)^2+\sigma_0^4\left(y^2-2\mu_1y+\mu_1^2\right)^2}{4\sigma_0^4\sigma_1^4}\right\} E_0\left\{\frac{\left(y^2-2\mu_0y+\mu_0^2\right)\left(y^2-2\mu_1y+\mu_1^2\right)}{2\sigma_0^2\sigma_1^2}\right\}\\
	&= E_0\left\{\frac{\left(y^4+4\mu_0^2y^2+\mu_0^4-4\mu_0y^3+2\mu_0^2y^2-4\mu_0^3y\right)}{4\sigma_0^4}\right\} + E_0\left\{\frac{\left(y^4+4\mu_1^2y^2+\mu_1^4-4\mu_1y^3+2\mu_1^2y^2-4\mu_1^3y\right)}{4\sigma_1^4}\right\}\\
	&\quad - E_0\left\{\frac{\left(y^4+4\mu_0\mu_1y^2+\mu_0^2\mu_1^2-2\mu_1y^3-2\mu_0y^3+\mu_0^2y^2+\mu_1^2y^2-2\mu_0\mu_1^2y-2\mu_0^2\mu_1y\right)}{2\sigma_0^2\sigma_1^2}\right\}\\
\end{align*}
\begin{align*}
	Z_3 &= \frac{\left(E_0\{y^4\}+4\mu_0^2E_0\{y^2\}+\mu_0^4-4\mu_0E_0\{y^3\}+2\mu_0^2E_0\{y^2\}-4\mu_0^3E_0\{y\}\right)}{4\sigma_0^4}\\
	&\quad + \frac{\left(E_0\{y^4\}+4\mu_1^2E_0\{y^2\}+\mu_1^4-4\mu_1E_0\{y^3\}+2\mu_1^2E_0\{y^2\}-4\mu_1^3E_0\{y\}\right)}{4\sigma_1^4}\\
	&\quad - \frac{\left(E_0\{y^4\}+4\mu_0\mu_1E_0\{y^2\}+\mu_0^2\mu_1^2-2\mu_1E_0\{y^3\}-2\mu_0E_0\{y^3\}+\mu_0^2E_0\{y^2\}+\mu_1^2E_0\{y^2\}-2\mu_0\mu_1^2E_0\{y\}-2\mu_0^2\mu_1E_0\{y\}\right)}{2\sigma_0^2\sigma_1^2}\\
	&=\frac{\mu_0^4+6\mu_0^2\sigma_0^2+3\sigma_0^4+6\mu_0^4+6\mu_0^2\sigma_0^2-4\mu_0^4-12\mu_0^2\sigma_0^2-4\mu_0^4+\mu_0^4}{4\sigma_0^4}\\&\quad + \frac{\mu_0^4+6\mu_0^2\sigma_0^2+3\sigma_0^4+6\mu_0^2\mu_1^2+6\mu_1^2\sigma_0^2-4\mu_0^3\mu_1-12\mu_0\mu_1\sigma_0^2-4\mu_0\mu_1^3+\mu_1^4}{4\sigma_1^4}\\&\quad - \frac{\mu_0^4+6\mu_0^2\sigma_0^2+3\sigma_0^4+\left(\mu_0^2+\sigma_0^2\right)\left(4\mu_0\mu_1+\mu_0^2+\mu_1^2\right)-\left(\mu_0^3+3\mu_0\sigma_0^2\right)\left(2\mu_0+2\mu_1\right)-\mu_0\left(2\mu_0\mu_1^2+2\mu_0^2\mu_1\right)+\mu_0^2\mu_1^2}{2\sigma_0^2\sigma_1^2}\\
	&= \frac{3}{4} + \frac{\mu_0^4+\mu_1^4-4\left(\mu_0^3\mu_1+\mu_0\mu_1^3\right)+6\left(\mu_0^2\sigma_0^2+\mu_1^2\sigma_0^2+\mu_0^2\mu_1^2-2\mu_0\mu_1\sigma_0^2\right)+3\sigma_0^4}{4\sigma_1^4} \\
	&\quad- \frac{\mu_0^4+6\mu_0^2\sigma_0^2+3\sigma_0^4+4\mu_0^3\mu_1+\mu_0^4+\mu_0^2\mu_1^2+4\mu_0\mu_1\sigma_0^2+\mu_0^2\sigma_0^2+\mu_1^2\sigma_0^2}{2\sigma_0^2\sigma_1^2}\\
	&\quad-\frac{-2\mu_0^4-2\mu_0^3\mu_1-6\mu_0^2\sigma_0^2-6\mu_0\mu_1\sigma_0^2-2\mu_0^2\mu_1^2-2\mu_0^3\mu_1+\mu_0^2\mu_1^2}{2\sigma_0^2\sigma_1^2}\\
	&=\frac{3}{4} + \frac{\mu_0^4+\mu_1^4-4\left(\mu_0^3\mu_1+\mu_0\mu_1^3\right)+6\left(\mu_0^2\sigma_0^2+\mu_1^2\sigma_0^2+\mu_0^2\mu_1^2-2\mu_0\mu_1\sigma_0^2\right)+3\sigma_0^4}{4\sigma_1^4}-\frac{3\sigma_0^2-2\mu_0\mu_1+\mu_0^2+\mu_1^2}{2\sigma_1^2}\\
\end{align*}
\begin{align*}
	\sigma_{\eta,0}^2 &= E_0\{\eta^2\} - \mu_{\eta,0}^2 
	= Z_3 + Z_2^2 + 2Z_1Z_2 - \left(Z_1^2+Z_2^2+2Z_1Z_2\right)\\
	&= Z_3 - \frac{\left(\mu_0^2+\mu_1^2-2\mu_0\mu_1+\sigma_0^2-\sigma_1^2\right)^2}{4\sigma_1^4}\\
	&=Z_3 - \frac{\mu_0^4+\mu_1^4+4\mu_0^2\mu_1^2+\sigma_0^4+\sigma_1^4+2\mu_0^2\mu_1^2-4\mu_0^3\mu_1+2\mu_0^2\sigma_0^2-2\mu_0^2\sigma_1^2-4\mu_0\mu_1^3}{4\sigma_1^4}\\
	&\quad-\frac{2\mu_1^2\sigma_0^2-2\mu_1^2\sigma_1^2-4\mu_0\mu_1\sigma_0^2+4\mu_0\mu_1\sigma_1^2-2\sigma_0^2\sigma_1^2}{4\sigma_1^4}\\
	&=\frac{3}{4} + \frac{4\sigma_0^2\left(\mu_0-\mu_1\right)^2 + 2\sigma_0^4-\sigma_1^4+2\sigma_1^2\left(\mu_0-\mu_1\right)^2+2\sigma_0^2\sigma_1^2}{4\sigma_1^4}-\frac{3\sigma_0^2+\left(\mu_0-\mu_1\right)^2}{2\sigma_1^2}\\
	&= \frac{1}{2} \left(1+ \frac{\sigma_0^4}{\sigma_1^4}\right) + \left(\mu_0-\mu_1\right)^2\frac{\sigma_0^2}{\sigma_1^4}-\frac{\sigma_0^2}{\sigma_1^2}
\end{align*}
The mean and the variance of the log-likelihood ratio under the alternative hypothesis are given by
\begin{align*}
	\mu_{\eta,1} &= \frac{\mu_0^2+\mu_1^2-2\mu_0\mu_1+\sigma_1^2-\sigma_0^2}{2\sigma_0^2}+ \log\!\left(\frac{\sigma_0}{\sigma_1}\right)\\
	\sigma_{\eta,1}^2 &=\frac{1}{2} \left(1+ \frac{\sigma_1^4}{\sigma_0^4}\right) + \left(\mu_0-\mu_1\right)^2\frac{\sigma_1^2}{\sigma_0^4}-\frac{\sigma_1^2}{\sigma_0^2}.
\end{align*}

\section{Decision Thresholds for the $\mathcal{CI}$SPRT}
\label{app:gen_th}

The probability of false alarm can be written as \cite{sahu2016distributed}

\begin{align*}
\begin{aligned}
	P_\text{FA} &= P_0(S_k(T) \geq \upsilon)
	\leq \sum_{t=1}^\infty P_0(S_k(t) \geq \upsilon)\\
	&\leq \sum_{t=1}^\infty \mathcal{Q}\left(\frac{\upsilon-\mu_{\eta,0}t}{\sigma_{\eta,0} \sqrt{\xi t}}\right).
	\end{aligned}
\end{align*}
Using the property $\mathcal{Q}(x) \leq \frac{1}{2}e^{-\frac{x^2}{2}}$ and following the derivation in \cite{sahu2016distributed}, we obtain
\begin{align*}
	P_\text{FA} &\leq \frac{1}{2}\sum_{t=1}^\infty e^{\frac{-\upsilon^2 - \mu_{\eta,0}^2t^2 + 2\upsilon\mu_{\eta,0}t}{2\sigma_{\eta,0}^2\xi t}}\\
	&= \frac{1}{2}e^{\frac{\upsilon\mu_{\eta,0}}{\sigma_{\eta,0}^2\xi}} \sum_{t=1}^\infty e^{\frac{-\upsilon^2 - \mu_{\eta,0}^2t^2}{2\sigma_{\eta,0}^2\xi t}}\\
	&= \frac{1}{2}e^{\frac{\upsilon\mu_{\eta,0}}{\sigma_{\eta,0}^2\xi}} \left[ \sum_{t=1}^{\lfloor \frac{\upsilon}{2\mu_{\eta,0}}\rfloor} e^{\frac{-\upsilon^2 - \mu_{\eta,0}^2t^2}{2\sigma_{\eta,0}^2\xi t}} 
	+ \sum_{t=\lfloor \frac{\upsilon}{2\mu_{\eta,0}}\rfloor+1}^{\lfloor \frac{\upsilon}{\mu_{\eta,0}}\rfloor} e^{\frac{-\upsilon^2 - \mu_{\eta,0}^2t^2}{2\sigma_{\eta,0}^2\xi t}} 
	+ \sum_{t=\lfloor \frac{\upsilon}{\mu_{\eta,0}}\rfloor+1}^{\lfloor \frac{2\upsilon}{\mu_{\eta,0}}\rfloor} e^{\frac{-\upsilon^2 - \mu_{\eta,0}^2t^2}{2\sigma_{\eta,0}^2\xi t}}  + \sum_{t=\lfloor \frac{2\upsilon}{\mu_{\eta,0}}\rfloor+1}^{\infty} e^{\frac{-\upsilon^2 - \mu_{\eta,0}^2t^2}{2\sigma_{\eta,0}^2\xi t}}\right]\\
	&\leq \frac{1}{2}e^{\frac{\upsilon\mu_{\eta,0}}{\sigma_{\eta,0}^2\xi}} 
	\left[ e^{-\frac{\upsilon\mu_{\eta,0}}{\sigma^2_{\eta,0}\xi}}\sum_{t=1}^{\lfloor \frac{\upsilon}{2\mu_{\eta,0}}\rfloor} e^{-\frac{\mu_{\eta,0}^2t}{2\sigma_{\eta,0}^2\xi}}
	+ e^{-\frac{\upsilon\mu_{\eta,0}}{2\sigma^2_{\eta,0}\xi}}\sum_{t=\lfloor \frac{\upsilon}{2\mu_{\eta,0}}\rfloor+1}^{\lfloor \frac{\upsilon}{\mu_{\eta,0}}\rfloor} e^{-\frac{\mu_{\eta,0}^2t}{2\sigma_{\eta,0}^2\xi}}
	 + e^{-\frac{\upsilon\mu_{\eta,0}}{4\sigma^2_{\eta,0}\xi}}\sum_{t=\lfloor \frac{\upsilon}{\mu_{\eta,0}}\rfloor+1}^{\lfloor \frac{2\upsilon}{\mu_{\eta,0}}\rfloor} e^{-\frac{\mu_{\eta,0}^2t}{2\sigma_{\eta,0}^2\xi}}  + \sum_{t=\lfloor \frac{2\upsilon}{\mu_{\eta,0}}\rfloor+1}^{\infty} e^{-\frac{\mu_{\eta,0}^2t}{2\sigma_{\eta,0}^2\xi}}\right].\\
	\end{align*}
	Approximating the sums above with infinite geometric series as in \cite{sahu2016distributed} and using the relation
	\begin{align*}
		\sum_{t=0}^{\infty}ar^t = \frac{a}{1-r},\qquad \text{for}\ |r| < 1,
	\end{align*}
	leads to an upper bound on the probability of false alarm according to
	\begin{align*}
	P_\text{FA}&\leq \frac{1}{2}\frac{e^{\frac{\upsilon\mu_{\eta,0}}{\sigma_{\eta,0}^2\xi}}}{1-e^{-\frac{\mu_{\eta,0}^2}{2\sigma_{\eta,0}^2\xi}}} \left[ e^{-\frac{\upsilon\mu_{\eta,0}}{\sigma^2_{\eta,0}\xi}}+e^{-\frac{\upsilon\mu_{\eta,0}}{2\sigma^2_{\eta,0}\xi}}e^{-\frac{\upsilon\mu_{\eta,0}}{4\sigma^2_{\eta,0}\xi}}
	 + e^{-\frac{\upsilon\mu_{\eta,0}}{4\sigma^2_{\eta,0}\xi}}e^{-\frac{\upsilon\mu_{\eta,0}}{2\sigma^2_{\eta,0}\xi}} + e^{-\frac{\upsilon\mu_{\eta,0}}{\sigma^2_{\eta,0}\xi}}\right]\\
	&\leq \frac{e^{\frac{4\upsilon\mu_{\eta,0}}{4\sigma_{\eta,0}^2\xi}}}{1-e^{-\frac{\mu_{\eta,0}^2}{2\sigma_{\eta,0}^2\xi}}} \left[ e^{-\frac{4\upsilon\mu_{\eta,0}}{4\sigma^2_{\eta,0}\xi}}+e^{-\frac{3\upsilon\mu_{\eta,0}}{4\sigma^2_{\eta,0}\xi}}\right]\\
	&\leq\frac{2e^{\frac{\upsilon\mu_{\eta,0}}{4\sigma_{\eta,0}^2\xi}}}{1-e^{-\frac{\mu_{\eta,0}^2}{2\sigma_{\eta,0}^2\xi}}}.
\end{align*}
Requiring $P_\text{FA} \leq \alpha$ and solving for $\upsilon$ yields the upper threshold $\upsilon$ as
\begin{align*}
	\alpha &\leq\frac{2e^{\frac{\upsilon\mu_{\eta,0}}{4\sigma_{\eta,0}^2\xi}}}{1-e^{-\frac{\mu_{\eta,0}^2}{2\sigma_{\eta,0}^2\xi}}}\\
	\frac{\alpha}{2}\left( 1-e^{-\frac{\mu_{\eta,0}^2}{2\sigma_{\eta,0}^2\xi}}\right) &\leq e^{\frac{\upsilon\mu_{\eta,0}}{4\sigma_{\eta,0}^2\xi}}\\
	\log\!\left(\frac{\alpha}{2}\right) + \log\!\left( 1-e^{-\frac{\mu_{\eta,0}^2}{2\sigma_{\eta,0}^2\xi}}\right) &\leq \frac{\upsilon\mu_{\eta,0}}{4\sigma_{\eta,0}^2\xi} \\
	\upsilon &\geq \frac{4\sigma_{\eta,0}^2\xi}{\mu_{\eta,0}} \left[ \log\!\left(\frac{\alpha}{2}\right) + \log\!\left( 1-e^{-\frac{\mu_{\eta,0}^2}{2\sigma_{\eta,0}^2\xi}}\right) \right].
	\label{eq:gamma_u}
\end{align*}
Repeating the same procedure for the probability of misdetection and requiring $P_\text{MD}\leq \beta$ yields the lower threshold
\begin{align*}
	\lambda &\leq \frac{4\sigma_{\eta,1}^2\xi}{\mu_{\eta,1}} \left[ \log\!\left(\frac{\beta}{2}\right) + \log\!\left( 1-e^{-\frac{\mu_{\eta,1}^2}{2\sigma_{\eta,1}^2\xi}}\right) \right].
\end{align*}

\twocolumn

\ifCLASSOPTIONcaptionsoff
  \newpage
\fi



%
\bibliographystyle{IEEEtran}
\bibliography{references.bib}

\end{document}